\documentclass[pra, reprint, twocolumn, superscriptaddress, amsfonts, amssymb, amsmath]{revtex4-2}
\usepackage[english]{babel}
\usepackage{bbm}
\usepackage{lipsum}
\usepackage{physics}
\usepackage{graphicx}
\usepackage{mathtools}
\usepackage{hyperref}
\usepackage{verbatim}
\usepackage{xcolor}
\usepackage[normalem]{ulem}

\newcommand*{\sump}{}
\DeclareRobustCommand*{\sump}{%
    \mathop{{\sum}^{\mathrlap{\prime}}}%
}

\begin{document}

\title{Self-consistent microscopic derivation of Markovian master equations\\
  for open quadratic quantum systems}

\date{\today}

\author{Antonio \surname{D'Abbruzzo}}
\affiliation{Dipartimento di Fisica dell'Università di Pisa, Largo Pontecorvo 3, I-56127 Pisa, Italy}

\author{Davide \surname{Rossini}}
\affiliation{Dipartimento di Fisica dell'Università di Pisa, Largo Pontecorvo 3, I-56127 Pisa, Italy}
\affiliation{INFN, Sezione di Pisa, Largo Pontecorvo 3, I-56127 Pisa, Italy}

\begin{abstract}
  We provide a rigorous construction of Markovian master equations for a wide class of
  quantum systems that encompass quadratic models of finite size,
  linearly coupled to an environment modeled by a set of independent thermal baths.
  Our theory can be applied for both fermionic and bosonic models in any number of physical
  dimensions, and does not require any particular spatial symmetry of the global system.
  We show that, for non-degenerate systems under a full secular approximation,
  the effective Lindblad operators
  are the normal modes of the system, with coupling constants that explicitly depend
  on the transformation matrices that diagonalize the Hamiltonian.
  Both the dynamics and the steady-state (guaranteed to be unique) properties
  can be obtained with a polynomial amount of resources in the system size.
  We also address the particle and energy current flowing through the system
  in a minimal two-bath scheme and find that they hold the structure of Landauer’s
  formula, being thermodynamically consistent.
\end{abstract}

\maketitle


\section{Introduction}
\label{sec:introduction}

In the past decade, the study of quantum many-body systems in contact with
some external environment has been receiving a great deal of attention,
in view of the amazing possibilities offered by a number of experimental platforms.
Atomic and molecular optical systems~\cite{Zoller-12}, as well as coupled QED
cavities~\cite{Houck-12, Carusotto-20} and optomechanical resonators~\cite{Marquardt-13},
to mention a few of them, enable us to achieve a remarkable degree of control
and readability in their microscopic components, so that genuine quantum
phenomena stemming from a nontrivial interplay of the coherent quantum dynamics
and dissipative effects may be carefully addressed in the near future.
Prototypical situations include the emergence of collective and critical
behaviors~\cite{Houck-14, Girvin-15, Fitzpatrick-17, Ma-19}, quantum transport
phenomena~\cite{Brantut-12, Krinner-15}, and quantum information processing based on
the generation and manipulation of entangled subsystems~\cite{Barreiro-11, Aolita-15}
or on quantum annealing~\cite{Dickson-13, Boixo-13, Mishra-18}.

From a theoretical point of view, addressing the many-body quantum dynamics
in a driven-dissipative context is considered a formidable task
and several approximations need to be invoked.
The modelization of an open quantum system itself poses delicate conceptual issues,
at the stage when the reduced dynamics of the system $\mathcal{S}$ under scrutiny
is posed in the form of a master equation~\cite{Petruccione, Rivas-12}.
Among the most commonly employed frameworks are the Caldeira-Leggett or the spin
boson model~\cite{Caldeira-81, Leggett-87}. The situation becomes more involved
if $\mathcal{S}$ is composed of many interacting subsystems, so that, depending on
the employed approximations, the resulting master equation may not even
preserve complete positivity of the density operator $\rho_{\mathcal S}(t)$,
as for the Redfield equations~\cite{Redfield-65}.
However it can be shown that, within the Markovian hypothesis, which holds provided
the bath relaxation time scales are much shorter than the time scales of interest
of the system dynamics, the time evolution of $\rho_{\mathcal S}(t)$ follows a
well-behaved master equation of the Lindblad-Gorini-Kossakowski-Sudarshan (LGKS)
type~\cite{Lindblad_original, GKS_original}.

Notwithstanding this approximation, it is rather intuitive to observe that,
if interactions among the various constituents of the system are properly taken into account,
the environment would introduce incoherent excitation mechanisms acting between
the different subsystems (see, e.g., Ref.~\cite{Petruccione}).
Then, the resulting master equation would require the knowledge of the
eigendecomposition of the system Hamiltonian $H_{\mathcal S}$, a task which is typically
hard to achieve, especially when the number of constituents increases.
For this reason, a vast majority of works in the many-body realm usually rely on heuristic
approaches and describe the effects of the environment on the system through local forms
of master equations: the common scenario is that of a LGKS master equation with Lindblad
jump operators acting locally in the physical space of the system (see, e.g.,
Refs.~\cite{Verstraete-04, Zwolak-04, Pizorn-08, Benenti-09,
  Diehl-10, Prosen-11, Lee-11, Cui-15, Werner-16, Jin-16,  Keck-17,
  FossFeig-17, Savona-19, Carleo-19, Ciuti-19, Yoshioka-19} and references therein).
It turns out that, for quantum optical implementations, the conditions leading to such
local approximation are typically satisfied~\cite{Zoller-12, Sieberer-16}, therefore this
formalism constitutes the standard choice for theoretical investigations of this kind of systems.

Unfortunately, the nonlocal terms neglected in the above mentioned treatment are crucial
to describe currents flowing into the system, as typically occurring in solid-state devices.
Indeed, local forms of master equations may lead to apparent thermodynamic inconsistencies,
as pointed out in Ref.~\cite{Kosloff}, or failure in grasping the critical behavior~\cite{Lutz-20}.
This spurred the quantum information community to investigate the
emerging differences between global and local master
equations~\cite{Rivas-10, Guimaraes-16, Volovich-16, Barranco-16, Dhar-16, Adesso-17,
  Hofer-17, Motz-17, Mitchinson-18, Tahir-18, Cattaneo-19, Mascarenhas-19} and to find
possible alternative schemes~\cite{Barra-15, Katz-16, Esposito-17, DeChiara-18, DeChiara-20, Farina-20}.

In this paper, we make a step forward in the treatment of open quantum many-body
systems and provide a full microscopic derivation of (nonlocal) LGKS master equation
for the wide class of quadratic models. Existing investigations of LGKS master equations
for quadratic many-body systems typically rely on a local system-environment
approach~\cite{Znidaric-10, Pizorn-08, Znidaric-10_b, Chatelain-17,
  Keck-17, Nigro-20, Rossini-20}. The possibility of having a nonlocal equation
has been considered much more rarely~\cite{Prosen-08, Horstmann-13} and a rigorous
microscopic derivation has been performed only for specific 
systems~\cite{Harbola, Cattaneo-20, Dorn-21, Santos, Benatti-20}.
A related class of quadratic bosonic system has been also studied exactly, i.e., without even
making the Born-Markov approximation~\cite{Martinez-13}. However, to our knowledge,
a statistics-independent formalism for generic quadratic systems is still lacking.

The method proposed here overcomes the limitations of local approaches by
making use of the spectrum of such systems, which represent one of the scarce,
yet paradigmatic, examples of exactly solvable quantum many-body systems.
Despite the fact that quadratic models cannot be deemed as truly interacting, being mappable into
free-quasiparticle systems, they are able to disclose a wealth of interesting phenomena
including topological phase transitions and critical behaviors~\cite{Kitaev-01, Peano-16}.

Our treatment goes through the diagonalization of $H_{\mathcal S}$,
which requires a number of resources scaling as twice the number $N$ of sites,
thus admitting to address systems with up to few thousands of sites.
In fact, it is possible to evaluate any kind of two-point observables
(as particle or energy currents) and of higher-order correlations
through the application of the Wick theorem.
We also stress that the analysis presented here works both for bosonic and for
fermionic particles, and is not restricted to any special geometry nor symmetry
in the system, being applicable to a variety of situations, which encompass existing
setups recently addressed in the literature~\cite{Santos, Benatti-20}.

The paper is organized as follows. In Sec.~\ref{sec:lindblad} we introduce the framework
we are going to focus on, which enables us to describe the temporal evolution of a quantum system
coupled to an external bath, under the weak-coupling, Born-Markov,
and secular approximations.
Section~\ref{sec:quadratic} contains a brief description of quadratic quantum many-body systems
and summarizes the general procedure that is needed to diagonalize them.
In Sec.~\ref{sec:rigorousME} we explicitly construct a class of Markovian master equations
for quadratic systems, following the self-consistent microscopic derivation outlined
in Sec.~\ref{sec:lindblad} which brings us to nonlocal dissipators.
Details on the procedure to obtain the temporal behavior and the asymptotics of two-point
observables and higher-order correlators are provided in Sec.~\ref{sec:steadystate},
where we also show that, for the master equation constructed in Sec.~\ref{sec:rigorousME},
the steady state is unique.
In Sec.~\ref{sec:twobath} we specialize to a minimal quantum-transport setup
composed of a one-dimensional system coupled to two baths at different temperatures
and chemical potentials. We discuss the possibility to establish steady-state particle
and energy currents, highlighting the emergence of thermoelectric effects and
showing the consistency with the thermodynamics, by proving the validity of the Onsager relation.
We conclude with a summary and perspectives for future work, in Sec.~\ref{sec:conclusions}.
Apps.~\ref{appendix} and~\ref{appendixB} discuss the subtleties that may emerge when the
system Hamiltonian supports zero-energy modes and/or degenerate eigenenergies, which
are cases that need to be treated separately.


\section{Markovian master equation}
\label{sec:lindblad}

We consider a quantum mechanical system $\mathcal{S}$ interacting with another
quantum system $\mathcal{E}$, that acts as an external environment.
By definition, the universe $\mathcal{U} = \mathcal{S} \, \cup \, \mathcal{E}$
(system plus environment) is a closed system and
the time evolution of its density operator $\rho_{\mathcal U}(t)$ is ruled by the Hamiltonian
\begin{equation}
  \label{eq:universe_hamiltonian}
  H_{\mathcal U} = H_\mathcal{S} \otimes I_\mathcal{E}
         + I_\mathcal{S} \otimes H_\mathcal{E} + H_{\rm int} \, ,
\end{equation}
where $H_\mathcal{S}$ ($H_\mathcal{E}$) denotes the free Hamiltonian of
$\mathcal{S}$ ($\mathcal{E}$), $I_\mathcal{S}$ ($I_\mathcal{E}$)
is the corresponding identity operator, and $H_{\rm int}$ is a term
describing the system-environment interaction.

The system's reduced density operator can be found by tracing out the environmental
degrees of freedom, through the identification
$\rho_\mathcal{S}(t) \equiv \Tr_\mathcal{E}[\rho_\mathcal{U}(t)]$.
Under the dynamical semigroup hypothesis,
such a reduction leads to the so-called LGKS Markovian master
equation~\cite{Petruccione, Rivas-12, Lindblad_original, GKS_original}
\begin{subequations}
  \label{eq:LGKS}
  \begin{equation}
    \frac{d\rho_\mathcal{S}(t)}{dt} =
    -i \big\{ H, \rho_\mathcal{S}(t) \big\}_ - + \mathcal{D}[\rho_\mathcal{S}(t)] \,,
  \end{equation}
  where $H$ is a Hermitian operator, generally differing from $H_\mathcal{S}$.
  The superoperator $\mathcal{D}[\cdot]$ is responsible for the dissipation
  and can be cast in the form
  \begin{equation}
    \mathcal{D}[\rho] = \sum_{i,j} a_{ij}
    \left( 2 L_i \rho L_j^\dagger - \big\{ L_j^\dagger L_i, \rho \big\}_+ \right),
  \end{equation}
where $a_{ij}$ are coupling constants and $L_j$ are the Lindblad operators.
\end{subequations}
The notation used above,
\begin{equation}
  \big\{ X, Y \big\}_\zeta \equiv X Y + \zeta Y X,
  \label{eq:commut}
\end{equation}
with $\zeta = \pm 1$, distinguishes between the anti-commutator ($\zeta=+1$)
and the commutator ($\zeta=-1$) of two operators.
For the sake of clarity in the notations, hereafter we will be working
in units of $\hbar = k_B = 1$.

Finding suitable expressions for the quantities entering Eqs.~\eqref{eq:LGKS}
from a given microscopic model can be a laborious problem, especially for complex systems.
In most occasions, phenomenologically-derived local system-bath coupling schemes
are assumed, so that typically the Lindblad operators $L_j$
act on an appropriate spatial coordinate (e.g., a single site of a quantum lattice model).
Despite the successes achieved in describing a variety of situations, as for quantum optical
devices~\cite{Sieberer-16}, it has been shown that such an approach can lead to contradictory
results, which may lead to a violation of the second principle of thermodynamics~\cite{Kosloff}.
For example, this can happen if different parts of $\mathcal{S}$ are strongly coupled
to each other, a fact which clearly hints at a breakdown of such a local approximation.

The flaw of this phenomenological approach resides in the lack of an appropriate derivation
process for the master equation from the microscopic dynamics.
The standard way to do that can be summarized
as follows~\cite{Petruccione}. Without loss of generality, one first needs to write
the spectral decomposition of the system Hamiltonian, $H_\mathcal{S} = \sum_k \omega_k \dyad{k}{k}$,
and the interaction Hamiltonian as $H_{\rm int} = \sum_\alpha O_\alpha \otimes R_\alpha$
(where $O_\alpha$ acts on $\mathcal{S}$ and $R_\alpha$ acts on $\mathcal{E}$).
This leads to the following:
\begin{subequations}
  \label{eq:LGKS-selfc}
   \begin{align}
     & H = H_\mathcal{S} + H_{LS} \equiv H_\mathcal{S} + \! \sum_{\alpha,\beta; \omega} S_{\alpha\beta}(\omega) O_\alpha^\dagger(\omega) O_\beta(\omega) , \label{eq:def_lambshift} \\
      & \mathcal{D}[\rho] \!= \!\!\! \sum_{\alpha,\beta; \omega} \! \Gamma_{\alpha\beta}(\omega) \Big[ 2O_\beta(\omega) \rho O_\alpha^\dagger(\omega) \! - \! \{ O_\alpha^\dagger(\omega) O_\beta(\omega), \rho \}_+ \! \Big] \! , \label{eq:microscopic_dissipator}
   \end{align}
\end{subequations}
where $H_{LS}$ is a Lamb-shift correction,
\begin{equation}
  \label{eq:def_eigenoperators}
   O_\alpha(\omega) \equiv \sum_{k,q} \delta_{\omega_q - \omega_k, \omega} \dyad{k}{k}\! O_\alpha \!\dyad{q}{q}
\end{equation}
are the eigenoperators of $H_\mathcal{S}$, and
\begin{align}
  \Gamma_{\alpha\beta}(\omega) &= \frac{1}{2} \int_{-\infty}^{\infty} \!\! d\tau \, e^{i\omega\tau} \langle \widetilde{R}_\alpha^\dagger(\tau) R_\beta \rangle \, , \label{eq:env_correlation}\\
  S_{\alpha\beta}(\omega) &= \frac{1}{2i} \int_{0}^{\infty} \!\! d\tau \left[ e^{i\omega\tau} \langle \widetilde{R}_\alpha^\dagger(\tau) R_\beta \rangle - e^{-i\omega\tau} \langle R_\alpha^\dagger \widetilde{R}_\beta(\tau) \rangle \right] \!. \label{eq:env_S}
\end{align}
In the above expressions, $\langle \cdot \rangle$ is the mean value calculated with the environmental
density operator $\rho_\mathcal{E}$, supposed to be constant by means of the Born-Markov hypothesis,
and $\widetilde{X}(\tau) \equiv e^{iH_\mathcal{E}\tau} X e^{-iH_\mathcal{E}\tau}$.
It is also important to highlight that Eqs.~\eqref{eq:LGKS-selfc} are obtained
after a full secular approximation, the validity of which is based on the assumption that the
system's eigenfrequencies $\{\omega_k\}$ are either degenerate or well-spaced with respect to the
system's typical evolution timescale~\cite{Petruccione}. In this work, we assume this condition
to be valid, thus excluding the presence of quasi-degeneracies. If they do occur, a more general
partial secular approximation needs to be invoked~\cite{Cattaneo-19}: without it, non-physical
results may emerge (as the absence of heat transport between different temperature
reservoirs---see Ref.~\cite{Wichterich-07}).
We leave this interesting issue to a future investigation.

The crucial point of Eqs.~\eqref{eq:LGKS-selfc} is the spectral decomposition of $H_\mathcal{S}$,
which is required to find the eigenoperators $O_\alpha(\omega)$.
In general, this is difficult to exploit and this is the reason why the microscopic derivation is
rarely used in physical situations, apart from specific cases of very simple quantum systems
(as, for example, a single quantum spin or a bunch of coupled qubits).
Below we apply this derivation to the class of quadratic quantum many-body systems which can be
effectively diagonalized with a polynomial amount of resources in the system size.


\section{Quadratic quantum systems}
\label{sec:quadratic}

In this section we summarize the basic properties of quadratic quantum systems, as
they constitute a paradigmatic example of many-body systems~\cite{Blaizot-Ripka},
focusing on the procedure that is needed to effectively obtain their spectral decomposition.
Consider a system $\mathcal{S}$ defined on a lattice with $N > 1$ sites,
and denote with $a_j^{(\dagger)}$ the annihilation (creation) operator associated with
the $j$th site, where $j=1,\ldots,N$. The set of these operators obeys the canonical
rules
\begin{equation}
  \label{eq:canonical_rules}
   \big\{ a_i, a_j^\dagger \big\}_\zeta = \delta_{ij} \, ,
   \qquad
   \big\{ a_i, a_j \big\}_\zeta = \big\{ a_i^\dagger, a_j^\dagger \big\}_\zeta = 0 \, ,
\end{equation}
where we have adopted the notation of Eq.~\eqref{eq:commut}, so 
$\zeta$ stores information about the statistics of the components of $\mathcal{S}$:
$\zeta = +1$ implies anti-commutation rules, holding for a fermionic system,
while $\zeta = -1$ is for commutation rules, holding for a bosonic system.

The most general free Hamiltonian of a quadratic system is given by
\begin{equation}
  \label{eq:quadratic_hamiltonian}
  H_\mathcal{S} = \sum_{i,j=1}^N \left[ Q_{ij} a_i^\dagger a_j
    + \tfrac{1}{2} \big( P_{ij} a_i^\dagger a_j^\dagger + P^*_{ij} a_j a_i \big) \right].
\end{equation}
The terms with coefficients $Q_{ij}$ are called normal terms, while those
with coefficients $P_{ij}$ are called anomalous (or pairing) terms,
since their presence makes $H_\mathcal{S}$ non number-conserving.
Note that the Hermiticity of $H_\mathcal{S}$ and the constraints
in Eq.~\eqref{eq:canonical_rules} impose the conditions
\begin{equation}
  \label{eq:constrQP}
    Q^\dagger = Q, \qquad P^T = -\zeta P
\end{equation}
on the coefficient matrices.
In this work, we assume for simplicity that $Q_{ij}$ and $P_{ij}$ are
time-independent coefficients. However, it is quite straightforward to generalize
our construction below to the case of time-dependent Hamiltonians.

The spectral decomposition of $H_\mathcal{S}$ can be obtained through a
Bogoliubov-Valatin (BV) transformation, which formulates the task in terms of
a standard linear-algebra eigenvalue problem.
To fix the notations, below we provide a brief description of it,
referring to~\cite{Blaizot-Ripka, vanHemmen, Xiao} for a more detailed discussion.

We first define the $2N$-dimensional Nambu field vector
\begin{equation}
  \mathbbm{a}^\dagger = (a_1^\dagger, \ldots, a_N^\dagger, a_1, \ldots, a_N),
\end{equation}
where blackboard bold letters denote objects living
in the doubled Nambu space on which the field vector acts.
The canonical rules~\eqref{eq:canonical_rules} translate into
\begin{equation}
  \big\{ \mathbbm{a}_\mu, \mathbbm{a}_\nu^\dagger \big\}_\zeta = \mathbb{I}^{(\zeta)}_{\mu\nu} \, ,
  \qquad
  \text{where  } \;
   \mathbb{I}^{(\zeta)} \equiv \left(
   \begin{array}{cc}
      I & 0 \\ 0 & \zeta I
   \end{array}
   \right),
\end{equation}
and $I$ is the $N \times N$ identity matrix. With these definitions,
the quadratic Hamiltonian~\eqref{eq:quadratic_hamiltonian} takes the compact form
\begin{equation}
  \label{eq:BdG}
  H_\mathcal{S} = \tfrac{1}{2} \big( \mathbbm{a}^\dagger \, \mathbb{H} \, \mathbbm{a}
  + \zeta \Tr Q \big) \, ,
\end{equation}
where
\begin{equation}
  \mathbb{H} \equiv \left(
  \begin{array}{cc}
    Q & P \\ -\zeta P^* & -\zeta Q^*
  \end{array}
  \right)
\end{equation}
is often referred to as the Bogoliubov-de Gennes Hamiltonian.
Note that Eq.~\eqref{eq:constrQP} implies that $\mathbb{H}$ is Hermitian, however the latter
can be seen as a coefficient matrix in the Nambu space and not as an actual Hamiltonian operator.

Let us now define the operators $\{b_k\}_{k=1,\ldots,N}$ through the canonical transformation
\begin{equation}
  \label{eq:BV}
   a_j = \sum_{k=1}^N \left( A_{jk}b_k + B_{jk}b_k^\dagger \right).
\end{equation}
In the Nambu space this can be written as
\begin{equation}
  \label{eq:BV-Nambu}
  \mathbbm{a} = \mathbb{T} \mathbbm{b} \, ,
  \qquad
  \mathbb{T} \equiv \left(
  \begin{array}{cc}
    A & B \\ B^* & A^*
  \end{array}
  \right).
\end{equation}
To preserve the canonical rules
$\big\{ \mathbbm{b}_\mu, \mathbbm{b}_\nu^\dagger \big\}_\zeta = \mathbb{I}^{(\zeta)}_{\mu\nu}$
on the operators $b_k$, we have to impose:
\begin{equation}
  \mathbb{I}^{(\zeta)}_{\mu\nu} = \big\{ \mathbbm{a}_\mu, \mathbbm{a}_\nu^\dagger \big\}_\zeta = \sum_{\sigma,\tau} \mathbb{T}_{\mu\sigma} \big\{ \mathbbm{b}_\sigma, \mathbbm{b}_\tau^\dagger \big\}_\zeta \mathbb{T}_{\tau\nu}^\dagger \,,
\end{equation}
leading us to the condition
\begin{equation}
  \label{eq:T_condition}
  \mathbb{T} \mathbb{I}^{(\zeta)} \mathbb{T}^\dagger = \mathbb{I}^{(\zeta)},
\end{equation}
or, in terms of the $A$ and $B$ matrices,
\begin{subequations}
  \label{eq:BV_constraints}
  \begin{eqnarray}
    A^\dagger A + \zeta B^T B^* = & AA^\dagger + \zeta BB^\dagger & = I \, , \\
    A^\dagger B + \zeta B^T A^* = & AB^T + \zeta BA^T & = 0 \, .
  \end{eqnarray}
\end{subequations}

Using the transformation~\eqref{eq:BV-Nambu}, we can write Eq.\eqref{eq:BdG} as
\begin{equation}
  H_\mathcal{S} = \tfrac{1}{2} \Big[ \mathbbm{b}^\dagger \big( \mathbb{I}^{(\zeta)} \mathbb{T}^{-1} \mathbb{D} \mathbb{T} \big) \mathbbm{b} + \zeta \Tr Q \Big] \, ,
\end{equation}
where we have used Eq.~\eqref{eq:T_condition} and defined $\mathbb{D} \equiv \mathbb{I}^{(\zeta)} \mathbb{H}$.
One can prove that~\cite{Blaizot-Ripka, vanHemmen, Xiao}, if $\mathbb{D}$ is diagonalizable
with real eigenvalues, then it is always possible to choose $\mathbb{T}$ in such a way
to obtain $\mathbb{T}^{-1} \mathbb{D} \mathbb{T} = \text{diag}(\omega_1, \ldots, \omega_N, -\omega_1, \ldots, -\omega_N)$,
where $\omega_j \geq 0$ (note that, if the matrix $\mathbb{D}$ has a null eigenvalue,
this always comes in pairs and thus has an even degeneracy).
In this case, expanding the Nambu representation,
\begin{eqnarray}
  \nonumber
  H_\mathcal{S} & = & \frac{1}{2} \sum_{k=1}^N \omega_k (b_k^\dagger b_k - \zeta b_k b_k^\dagger) + \frac{\zeta}{2} \Tr Q \\
  & = & \sum_{k=1}^N \omega_k b_k^\dagger b_k + \frac{\zeta}{2} \bigg[ \Tr Q - \sum_{k=1}^N \omega_k \bigg] \, ,
\label{eq:diagonalized_hamiltonian}
\end{eqnarray}
where we have used $b_k b_k^\dagger = 1-\zeta b_k^\dagger b_k$ and $\zeta^2 = 1$.
This is the diagonalized form of the Hamiltonian $H_\mathcal{S}$: the set $\{\omega_k\}$
is the spectrum of excitations and $b_k$ assumes the role of the annihilation operator
of a normal mode (or quasiparticle excitation) with energy $\omega_k$.
If $\mathbb{T}$ can be chosen in this way, Eq.~\eqref{eq:BV} is called BV transformation
and the matrices $A,B$ are the BV matrices.
Note that for fermionic systems $(\zeta = +1)$ it is always possible to perform
such a transformation, since $\mathbb{D} = \mathbb{H}$ is Hermitian, and hence always
diagonalizable with real eigenvalues.
In contrast, for bosonic systems $(\zeta = -1)$ this is not always the case, nonetheless
it can be shown that if $H_\mathcal{S}$ is stable (i.e., $\mathbb{H}$ is positive definite)
then $\mathbb{D}$ has real positive eigenvalues $\omega_j > 0$ and the BV transformation
can be performed~\cite{Blaizot-Ripka, vanHemmen, Xiao}.
Situations where zero-energy bosonic modes (also known as soft modes) are present
are trickier to handle, as they can require a special type of diagonalization~\cite{Colpa},
which we do not discuss here.
From a numerical point of view, note that the problem of finding $\mathbb{T}$ is equivalent
in complexity to the diagonalization of $\mathbb{D}$, which is a matrix of size $2N \times 2N$.

Let us finally address the special case of a normal system, in which all the anomalous
terms are absent, i.e., $P_{ij}=0$ in Eq.~\eqref{eq:quadratic_hamiltonian}.
In such case, the free Hamiltonian can be simply written
as $H_\mathcal{S} = \textbf{a}^\dagger Q \, \textbf{a}$,
after defining the $N$-dimensional field vector $\textbf{a}^\dagger = (a_1^\dagger, \ldots, a_N^\dagger)$.
Thus, the problem of diagonalizing $\mathbb{D}$ translates into that of diagonalizing
the Hermitian matrix $Q$. Given the unitary matrix $A$ which diagonalizes it,
the transformation $\textbf{a} = A \, \textbf{b}$ is able to solve the problem, since
\begin{equation}
   H_\mathcal{S} = \textbf{b}^\dagger (A^\dagger Q A) \textbf{b} = \sum_{k=1}^N \omega_k b_k^\dagger b_k \, ,
\end{equation}
where $\omega_k$ are the eigenvalues of $Q$.
This coincides with a BV transformation with $B=0$ (i.e., there is no mixing between
annihilation and creation operators); in that case, the constraints~\eqref{eq:BV_constraints}
guarantee that $A$ is a unitary matrix and thus the total number of particles
$\mathcal{N} \equiv \sum_i a_i^\dagger a_i$ coincides with the total number of quasiparticles
$\mathcal{N}_Q \equiv \sum_k b_k^\dagger b_k$.


\section{Construction of the master equation for quadratic systems}
\label{sec:rigorousME}

In this section we show how to explicitly derive a realistic LGKS master equation for quadratic systems,
following the microscopic derivation outlined in Sec.~\ref{sec:lindblad} and the diagonalization procedure
described in Sec.~\ref{sec:quadratic}. This will lead to a nonlocal system-bath coupling,
which can nevertheless be handled within the BV formalism.


\subsection{Definition of the universe Hamiltonian}

The first step consists in the specification of the universe Hamiltonian (\ref{eq:universe_hamiltonian})
for the quadratic model in Eq.~(\ref{eq:quadratic_hamiltonian}).
We suppose that the environment consists of a set of $N_B$ independent thermal baths,
indexed by $n \in \{ 1,\ldots,N_B \}$, each of them characterized by a temperature $T_n$
and a chemical potential $\mu_n$. It is reasonable to assume that they are all described
by a continuous free model, such that
\begin{equation}
   H_\mathcal{E} = \sum_{n=1}^{N_B} \int dk \, \epsilon_n(k) \, c_n^\dagger(k) c_n(k) \equiv \sum_{n=1}^{N_B} H_{\mathcal{E},n} \, .
\end{equation}
where the spectrum $\epsilon_n(k) \geq 0$ is assumed to be non-negative.
The operators $c_n(k)$ fulfill the canonical rules
\begin{subequations}\label{eq:canonical_rules_bath}
\begin{eqnarray}
   \big\{ c_n(k), c_m^\dagger(q) \big\}_\zeta & = & \delta_{nm} \, \delta(k-q) \, , \\
   \big\{ c_n(k), c_m(q) \big\}_\zeta & = & \big\{ c_n^\dagger(k), c_m^\dagger(q) \big\}_\zeta = 0 \, ,
\end{eqnarray}
\end{subequations}
and satisfy the following relations (for any $\zeta = \pm 1$):
\begin{subequations}
   \label{eq:env_commproperty}
   \begin{eqnarray}
      \big\{ H_\mathcal{E}, c_n(k) \big\}_- & = & -\epsilon_n(k) \, c_n(k) \, , \\
      \big\{ H_\mathcal{E}, c_n^\dagger(k) \big\}_- & = & \epsilon_n(k) \, c_n^\dagger(k) \, .
   \end{eqnarray}
\end{subequations}
Moreover, by the hypothesis of independent thermal baths, the environmental
reduced density operator $\rho_\mathcal{E}$ assumes the factorized form
\begin{equation}
  \label{eq:env_densityoperator}
  \rho_\mathcal{E} = \bigotimes_{n=1}^{N_B} \frac{ e^{-(H_{\mathcal{E},n} - \mu_n \mathcal{N}_{\mathcal{E},n})/T_n} }
        { \Tr \big[ e^{-(H_{\mathcal{E},n} - \mu_n \mathcal{N}_{\mathcal{E},n})/T_n} \big] } \, ,
\end{equation}
where $\mathcal{N}_{\mathcal{E},n} \equiv \int dk \, c_n^\dagger(k) c_n(k)$.
From this, one can easily obtain the two-point expectation values
\begin{subequations}
  \label{eq:env_expvalues}
  \begin{align}
    & \big\langle c_n(k) \, c_m(q) \big\rangle = \langle c_n^\dagger(k) \, c_m^\dagger(q) \rangle = 0 \, , \\
    & \big\langle c_n^\dagger(k) \, c_m(q) \big\rangle = \delta_{nm} \, \delta(k-q) \, f_n(\epsilon_n(k)) \, ,
  \end{align}
\end{subequations}
with
\begin{equation}
  f_n(\epsilon) \equiv \left[ \zeta + e^{(\epsilon-\mu_n)/T_n} \right]^{-1} \, ,
  \label{eq:bath_statistics}
\end{equation}
being either the Fermi-Dirac distribution (if $\zeta = +1$) or the Bose-Einstein
distribution (if $\zeta = -1$).
We note that hereafter we always assume $\zeta=+1$ (for fermions)
or $\zeta=-1$ (for bosons),
both in Eqs.~\eqref{eq:canonical_rules}, \eqref{eq:canonical_rules_bath},
and~\eqref{eq:bath_statistics},
thus ignoring the cases where the system and the bath components obey different statistics.
The latter, mixed, case can however be easily taken into account using the same formalism.

\begin{figure}
   \includegraphics[width=0.9\columnwidth]{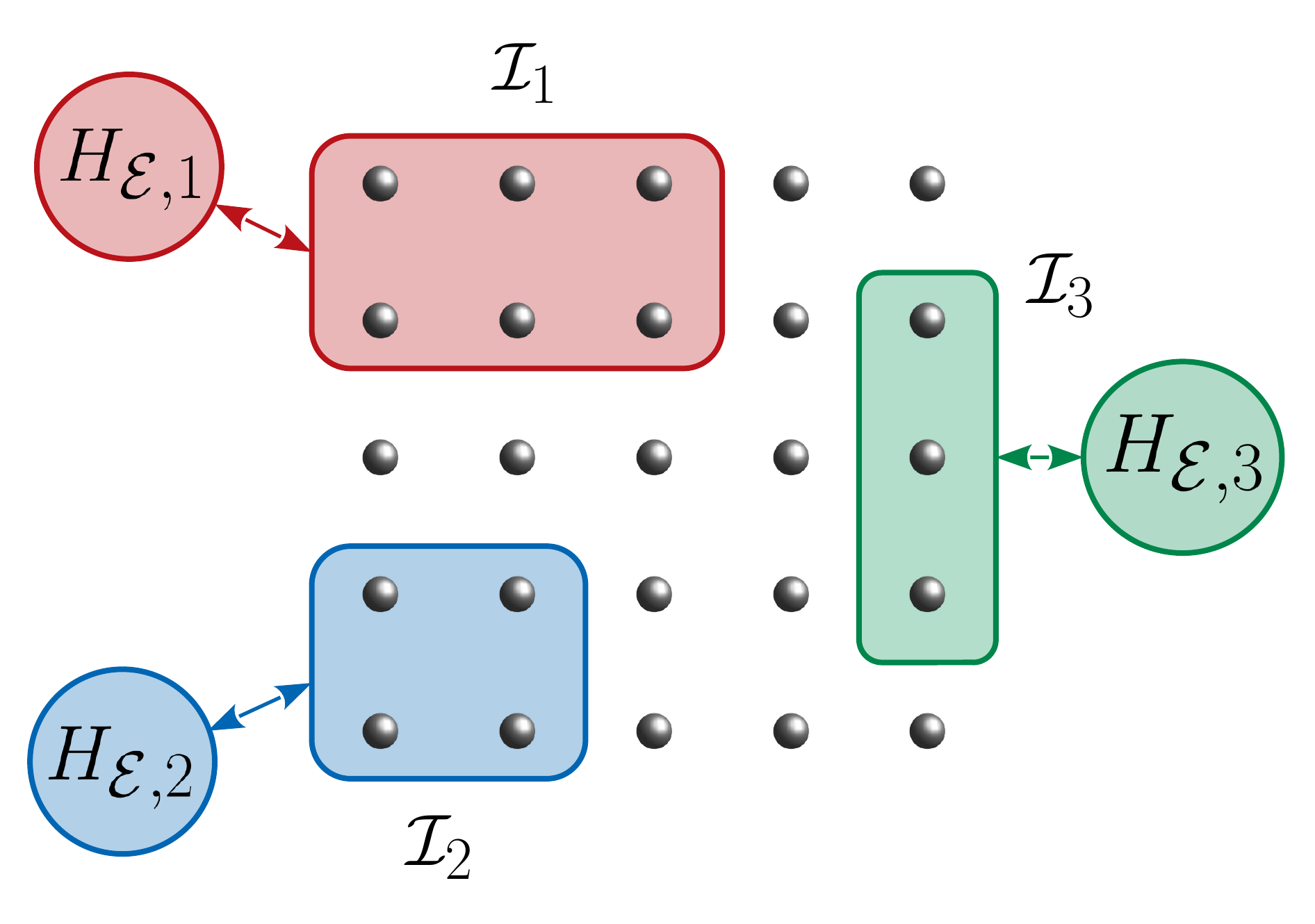}
   \caption{Schematic picture of system-environment interaction setting with $N_B = 3$,
     as described by Eq.~\eqref{eq:interaction}. Gray dots stand for the lattice sites
     of the system $\mathcal{S}$, while colored circles denote the external baths
     composing the environment $\mathcal{E}$.
     The highlighted regions in the lattice denote the sets of sites $\mathcal{I}_n$
     coupled to the $n$th bath.
     Note that the sketch is not in scale, since real baths are typically
     much larger than the system.}
   \label{fig:setting}
\end{figure}

As for the system-environment interaction, we consider a general linear coupling
between the environment variables and the sites of the system, that is:
\begin{equation}
  \label{eq:interaction}
  H_{\rm int} \!= \sum_{n=1}^{N_B} \sum_{p \in \mathcal{I}_n} \! \int \! dk \, g_{n}(k) w_{p,n}
  \left(a_p + a_p^\dagger \right) \left[c_n(k) + c_n^\dagger(k) \right] ,
\end{equation}
where the index $p$ runs on the set $\mathcal{I}_n$ gathering the lattice sites which are
physically coupled to the $n$th bath: this allows to consider various kinds
of interactions, even inhomogeneous ones (see the sketch in Fig.~\ref{fig:setting}).
The complex coefficient $g_{n}(k)$ quantifies the interaction strength between
the $k$th mode of the $n$th bath and the system
[in the Markovian hypothesis, it is reasonable to assume that $g_n(k)$ is uniform
  over the system sites to which the $k$th mode of the $n$th bath is coupled].
The coefficient $w_{p,n}$ is a site-dependent weight which can be used
to take into account, for instance, inhomogeneous spatial distributions of the couplings to a common environment.
The interaction Hamiltonian can also be written in the canonical form
$H_{\rm int} = \sum_{n=1}^{N_B} O_n \otimes R_n$, where
\begin{subequations}
  \label{eq:OandR}
   \begin{eqnarray}
      O_n & = & \sum_{p\in\mathcal{I}_n} w_{p,n} \left(a_p + a_p^\dagger \right) \, ,\\
      \label{eq:R}
      R_{n} & = & \int dk \, g_{n}(k) \left[ c_n(k) + c_n^\dagger(k) \right] \, .
   \end{eqnarray}
\end{subequations}


\subsection{Eigenoperators of the system Hamiltonian}

Now, suppose $\{ \ket{\textbf{x}} = \ket{x_1, \ldots, x_N} \}$ is the orthonormal basis
of the diagonalized quadratic hamiltonian~\eqref{eq:diagonalized_hamiltonian},
where $x_k \in \mathbb{N}$ is the occupation number associated with the $k$th normal mode
($x_k\,\in\,\{0,1\}$ for fermionic systems,
while $x_k\,\in\,\{0,1,2, \ldots\}$ for bosonic systems).
The energy of $\ket{\textbf{x}}$ is given by $E(\textbf{x}) \equiv \sum_{k=1}^N x_k \omega_k$.
With these notations, the definition of eigenoperator~\eqref{eq:def_eigenoperators}
associated with $O_n$ becomes
\begin{equation}
  O_n(\omega) = \sum_{p\in\mathcal{I}_n} w_{p,n} \sum_{\textbf{x}, \textbf{y}} \delta_{E(\textbf{y}) - E(\textbf{x}), \omega} \dyad{\textbf{x}}{\textbf{x}} \big( a_p + a^\dagger_p \big) \dyad{\textbf{y}}{\textbf{y}} \, .
  \label{eq:eigenop}
\end{equation}
Using the BV transformation~\eqref{eq:BV}, we can see that
\begin{equation}
   \mel{\textbf{x}}{a_p}{\textbf{y}} = \sum_{k=1}^N \left( A_{pk} \mel{\textbf{x}}{b_k}{\textbf{y}} + B_{pk} \mel{\textbf{x}}{b_k^\dagger}{\textbf{y}} \right) \, .
\end{equation}
The states $\ket{\textbf{x}}$ and $\ket{\textbf{y}}$ must be equal, except for their value
at the $k$th position, in order to have a non-zero expression.
In particular, the matrix element $\mel{\textbf{x}}{b_k}{\textbf{y}}$ is non-zero if
and only if $\, b_k \! \ket{\textbf{y}} = \ket{\textbf{x}}$, which implies
\begin{subequations}
   \begin{equation}
     \dyad{\textbf{x}}{\textbf{y}} = b_k \, , \quad
      E(\textbf{y}) - E(\textbf{x}) = \omega_k \, ,
   \end{equation}
   while a non-zero value of $\mel{\textbf{x}}{b_k^\dagger}{\textbf{y}}$ implies
   \begin{equation}
     \hspace*{11mm} \dyad{\textbf{x}}{\textbf{y}} = b_k^\dagger \, , \quad
     E(\textbf{y}) - E(\textbf{x}) = -\omega_k \, .
   \end{equation}
\end{subequations}
We can then write:
\begin{multline}
  \label{eq:middle_eigen1}
  \sum_{\textbf{x}, \textbf{y}} \delta_{E(\textbf{y}) - E(\textbf{x}), \omega} \dyad{\textbf{x}}{\textbf{x}} \! a_p \! \dyad{\textbf{y}}{\textbf{y}} \\
  \hspace{2cm} = \sum_k \big( A_{pk}\delta_{\omega, \omega_k} b_k + B_{pk} \delta_{\omega,-\omega_k} b_k^\dagger \big) \, ,
\end{multline}
and similarly, for $a_p^\dagger$,
\begin{multline}
  \label{eq:middle_eigen2}
  \sum_{\textbf{x}, \textbf{y}} \delta_{E(\textbf{y}) - E(\textbf{x}), \omega} \dyad{\textbf{x}}{\textbf{x}} \! a_p^\dagger \! \dyad{\textbf{y}}{\textbf{y}} \\
  \hspace{2cm} = \sum_k \big( B^*_{pk}\delta_{\omega, \omega_k} b_k + A^*_{pk} \delta_{\omega,-\omega_k} b_k^\dagger \big) \, .
\end{multline}
Adding Eq.~\eqref{eq:middle_eigen1} with~\eqref{eq:middle_eigen2}, we finally obtain
the complete set of eigenoperators of $H_\mathcal{S}$ as defined in Eq.~\eqref{eq:eigenop}:
\begin{equation}
  \label{eq:eigenoperators}
  O_n(\omega) = \sum_{p\in\mathcal{I}_n} w_{p,n} \sum_{k=1}^N \left[ \phi_{pk} \delta_{\omega, \omega_k} b_k
    + \phi^*_{pk} \delta_{\omega, -\omega_k} b_k^\dagger \right] \, ,
\end{equation}
where we introduced the matrix
\begin{equation}
   \phi \equiv A + B^* \,,
\end{equation}
for convenience of notation.

It is important to stress that, although the interaction operator $O_n$ has a local shape
(it acts only on $\mathcal{I}_n$), the corresponding eigenoperator is intrinsically
nonlocal, since it is composed of delocalized excitation operators.
We recover a local shape of $O_n(\omega)$ only after assuming that there is no coupling between
the different sites of $\mathcal{S}$. For example, taking a normal Hamiltonian with $B=0$ and
assuming $\omega_k \simeq \Omega, \; \forall k$, the Kronecker deltas in Eq.~\eqref{eq:eigenoperators}
can be pulled out of the sum over $k$ to obtain
\begin{equation}
  \label{eq:local_eigenoperators}
  O_n(\omega) \simeq \sum_{p\in\mathcal{I}_n} w_{p,n}
  \left[ \delta_{\omega, \Omega} \, a_p + \delta_{\omega, -\Omega} \, a_p^\dagger \right]  \, ,
\end{equation}
and this would have a local shape. Intuitively, this is equivalent to saying that the $a_p$
themselves are the normal modes of the system. For general quadratic Hamiltonians,
there is no reason to assume that the local approximation is valid, so to obtain physically
consistent results, one would necessarily have to stick with Eq.~\eqref{eq:eigenoperators}.

The next step should be to put the expression~\eqref{eq:eigenoperators} we obtained
for $O_n(\omega)$ into the microscopic dissipator of Eq.~\eqref{eq:microscopic_dissipator}.
Notice that, in our case, the Greek indexes $\alpha$ and $\beta$ should be replaced
with $n$ and $m$.
Before doing that, it is convenient to calculate explicitly the Fourier transform $\Gamma_{nm}(\omega)$
of the environment correlation functions, defined in Eq.~\eqref{eq:env_correlation}.


\subsection{Environment correlation functions}

Let us start by using Eq.~\eqref{eq:R} to write:
\begin{multline}
  \langle \widetilde{R}_n^\dagger(\tau) R_m \rangle = \int \! dk \int \! dq \: g^*_n(k) \, g_m(q) \\
  \hspace*{1mm} \times\left\langle e^{iH_\mathcal{E}\tau} \! \left[ c_n(k) + c_n^\dagger(k) \right] \!
  e^{-iH_\mathcal{E}\tau} \! \left[ c_m(q) + c_m^\dagger(q) \right] \right\rangle \! .
  \label{eq:envcorr1}
\end{multline}
This expression can be simplified by means of the Baker-Campbell-Hausdorff formula,
according to which, given two generic operators $X$ and $Y$,
\begin{equation}
  e^X Y e^{-X} = e^v Y \, , \qquad
   \text{if } \: \big\{ X,Y \big\}_- = vY, \, v \in \mathbb{C} \, .
\end{equation}
Remembering that the operators $c_n(k)$ must satisfy Eq.~\eqref{eq:env_commproperty},
the second line in Eq.~\eqref{eq:envcorr1} can be written as
\begin{multline}
  e^{-i\epsilon_n(k)\tau} \Big\langle c_n(k) \big[ c_m(q) + c_m^\dagger(q) \big] \Big\rangle \\
  +e^{i\epsilon_n(k)\tau} \Big\langle c_n^\dagger(k) \big[ c_m(q) + c_m^\dagger(q) \big] \Big\rangle \, .
\end{multline}
Due to Eqs.~\eqref{eq:env_expvalues} such expectation values can be explicitly calculated as
\begin{equation}
  \delta_{nm} \! \left\{ \! e^{-i\epsilon_n(k)\tau} \big[ 1 \! - \! \zeta f_n(\epsilon_n(k)) \big]
  \!+ \! e^{i\epsilon_n(k)\tau} f_n(\epsilon_n(k)) \! \right\} \!.
\end{equation}
The only non-zero values for the environment correlation functions occur when $n=m$,
therefore the only relevant term reads
\begin{multline}
  \label{eq:RR}
  \langle \widetilde{R}_n^\dagger(\tau) R_n \rangle = \int dk \, | g_n(k) |^2 \\
   \times \! \left\{ e^{i\epsilon_n(k)\tau} f_n(\epsilon_n(k))
  + e^{-i\epsilon_n(k)\tau} \big[ 1 \! - \! \zeta f_n(\epsilon_n(k)) \big] \right\} \!.
\end{multline}
The Fourier transform of $e^{\pm i \epsilon \tau}$ is $2\pi\delta(\omega \pm \epsilon)$, therefore
\begin{multline}
  \label{eq:middle_gamma}
  \Gamma_{nn}(\omega) = \pi \int dk \, | g_n(k) |^2 \\
  \times \Big\{ \! \delta(\omega + \epsilon_n(k)) f_n(\epsilon_n(k)) + \delta(\omega - \epsilon_n(k))
  \big[ 1-\zeta f_n(\epsilon_n(k)) \big] \! \Big\} \,.
\end{multline}

It is now convenient to define the spectral density associated with the $n$th bath as
\begin{equation}
  \label{eq:DOS}
   \mathcal{J}_n(\omega) \equiv \pi \int dk \, |g_n(k)|^2 \, \delta(\omega - \epsilon_n(k)) \, .
\end{equation}
Since $\epsilon_n(k) \geq 0$, we have $\mathcal{J}_n(\omega) = 0$ for $\omega < 0$.
In this way, Eq.~\eqref{eq:middle_gamma} can be rewritten as:
\begin{equation}
  \label{eq:gamma}
   \Gamma_{nn}(\omega) = \makeatletter \bBigg@{4}\{ \makeatother
   \begin{array}{ll}
      \mathcal{J}_n(\omega) \big[ 1-\zeta f_n(\omega) \big] & \text{if } \omega > 0 \,, \vspace*{1mm}\\
      \mathcal{J}_n(-\omega)f_n(-\omega) & \text{if } \omega < 0 \,, \vspace*{1mm}\\
      \mathcal{J}_n(0) \big[ 1+(1-\zeta) f_n(0) \big] & \text{if } \omega = 0 \,.
   \end{array}
\end{equation}


\subsection{Calculation of the dissipator}\label{subsec:dissipator}
In the previous subsection we showed that the matrix $\Gamma(\omega)$ is diagonal
in the bath index $n$. This means that, in our case, Eq.~\eqref{eq:microscopic_dissipator}
acquires the diagonal form
\begin{equation}
  \label{eq:our_dissipator_diagonal}
  \mathcal{D}[\rho] = \! \sum_{n; \omega} \Gamma_{nn}(\omega)
  \Big[ 2O_n(\omega) \rho O_n^\dagger(\omega) - \big\{ O_n^\dagger(\omega) O_n(\omega), \rho \big\}_{\!+} \Big] .
\end{equation}
We can now plug in the expression for $O_n(\omega)$ reported in Eq.~\eqref{eq:eigenoperators}.
Let us first look at the term $O_n(\omega)\rho \, O_n^\dagger(\omega)$:
\begin{align}
  \label{eq:OnOn}
  \!\!\! & O_n(\omega) \rho \, O_n^\dagger(\omega) = \nonumber \\
  \!\! & \sum_{p,s\in\mathcal{I}_n} w_{p,n} \, w^*_{s,n} \sum_{k,q=1}^N
  \! \left[ \phi_{pk} \, \delta_{\omega,\omega_k} \, b_k + \phi^*_{pk} \, \delta_{\omega,-\omega_k} \, b_k^\dagger \right] \nonumber \\
  & \times \rho \Big[ \phi^*_{sq} \, \delta_{\omega,\omega_q} \, b_q^\dagger
    + \phi_{sq} \, \delta_{\omega,-\omega_q} \, b_q \Big] \!.
\end{align}
For the sake of simplicity, let us now suppose that the system $\mathcal{S}$ does not have
degenerate eigenenergies and that it does not support a zero-energy mode, which means that
$\omega_k = \omega_q$ only if $k=q$ and there is no $k$ such that $\omega_k = 0$.
If this is the case, then the sum over $k,q$ in Eq.~\eqref{eq:OnOn} reduces to
\begin{equation}
  \label{eq:OnOn2}
  \sum_{k=1}^N \left[ \phi_{pk}\phi^*_{sk} \delta_{\omega,\omega_k} b_k \rho b_k^\dagger
    + \phi^*_{pk}\phi_{sk} \delta_{\omega,-\omega_k} b_k^\dagger \rho b_k \right] .
\end{equation}
The same simplification can be performed on the other terms of Eq.~\eqref{eq:our_dissipator_diagonal}.
The computation is quite straightforward and the result is
\begin{multline}\label{eq:middle_dissipator}
  \mathcal{D}[\rho] = \sum_{n;\, k} \Phi_{n,k}
  \Big[ \Gamma_{nn}(\omega_k) \left( 2b_k\rho b_k^\dagger - \big\{ b_k^\dagger b_k, \rho \big\}_+ \right) \\
    + \Gamma_{nn}(-\omega_k) \left( 2b_k^\dagger \rho b_k - \big\{ b_k b_k^\dagger, \rho \big\}_+ \right) \Big] \,,
\end{multline}
where
\begin{equation}\label{eq:Phi}
  \Phi_{n,k} \equiv \sum_{p,s\in\mathcal{I}_n} w_{p,n} \, w^*_{s,n} \, \phi_{pk} \,\phi^*_{sk}
  = \Big| \sum_{p\in\mathcal{I}_n} w_{p,n} \, \phi_{pk} \Big|^2 \geq 0\,.
\end{equation}
Note that the sum over $\omega$ has been performed taking advantage of the Kronecker deltas.
For the sake of compactness in the notations, hereafter we will always
implicitly assume that the index $k$ runs from $1$ to $N$, the bath index $n$ runs
from $1$ to $N_B$, while the index $p$ runs in $\mathcal{I}_n$.

We can now use Eq.~\eqref{eq:gamma} to finally obtain
\begin{multline}
  \label{eq:dissipator}
   \mathcal{D}[\rho] = \sum_{n; k} \gamma_{n,k} \Big[ \big( 1-\zeta f_n(\omega_k) \big)
   \Big( 2b_k\rho b_k^\dagger - \big\{ b_k^\dagger b_k, \rho \big\}_+ \Big) \\
   + f_n(\omega_k) \left( 2b_k^\dagger \rho b_k - \big\{ b_k b_k^\dagger, \rho \big\}_+ \right) \Big] \,,
\end{multline}
where we have introduced the coupling constants
\begin{equation}\label{eq:coupling}
   \gamma_{n,k} \equiv \mathcal{J}_n(\omega_k) \, \Phi_{n,k} \geq 0 \,.
\end{equation}
We emphasize that this dissipator is valid as long as $\omega_k \neq 0$ for all $k$.
If the system $\mathcal{S}$ supports a zero-energy mode, one can nevertheless follow the same
kind of procedure, but special care must be taken when manipulating products of eigenoperators,
as in Eq.~\eqref{eq:OnOn}. We defer a discussion of this case to App.~\ref{appendix}.
Attention should be also paid if the system supports degenerate eigenenergies:
the reader can find details on this issue in App.~\ref{appendixB}.

It is worth mentioning that, if one would have chosen the local approximate version
for the eigenoperators~\eqref{eq:local_eigenoperators}, the dissipator would have taken the form:
\begin{align}
  \nonumber \mathcal{D}^{(\text{l})}[\rho] \!= \! \sum_{n;p} \! \mathcal{J}_n(\Omega)
  & \Big[ (1-\zeta f_n(\Omega)) \Big( 2a_p \rho a_p^\dagger - \big\{ a_p^\dagger a_p, \rho \big\}_{\!+} \Big) \\
    & + \! f_n(\Omega) \Big( 2a_p^\dagger \rho a_p \!- \! \big\{ a_p a_p^\dagger, \rho \big\}_{\!+} \Big) \Big]
\end{align}
which is basically the usual local dissipator,
where $\gamma_p^{(\uparrow)} \equiv \sum_n \mathcal{J}_n(\Omega) f_n(\Omega)$ and
$\gamma_p^{(\downarrow)} \equiv \sum_n \mathcal{J}_n(\Omega) [1-\zeta f_n(\Omega)]$
quantify the population and depopulation rates of the $p$th
site~\cite{Pizorn-08, Benenti-09, Znidaric-10, Znidaric-10_b, Prosen-11, Chatelain-17, Keck-17}.


\subsection{Lamb-shift correction}

To conclude the construction, we have to calculate the Lamb-shift correction~\eqref{eq:def_lambshift}
to the free system Hamiltonian.
In the definition of the matrix $S(\omega)$ reported in Eq.~\eqref{eq:env_S}, the
environment correlation functions appear and the same argument as before applies,
therefore only diagonal terms with $n=m$ remain. This implies that
\begin{equation}
   H_{LS} = \sum_{n;\, \omega} S_{nn}(\omega) \, O_n^\dagger(\omega) \, O_n(\omega) \,.
\end{equation}
Inserting Eq.~\eqref{eq:eigenoperators}, we get
\begin{equation}\label{eq:middle_lambshift_pr}
   H_{LS} = \sum_{n;\, k} \Phi_{n,k} \left[ S_{nn}(\omega_k) \, b_k^\dagger b_k + S_{nn}(-\omega_k) \, b_k b_k^\dagger \right] \,.
\end{equation}
Neglecting a constant which will not appear in the master equation,
since $H_{LS}$ only enters via a commutator, we can safely rewrite
\begin{equation}
  \label{eq:middle_lambshift}
   H_{LS} = \sum_{n;\, k} \Phi_{n,k} \Big[ S_{nn}(\omega_k) - \zeta S_{nn}(-\omega_k) \Big] b_k^\dagger b_k \,.
\end{equation}
To proceed further, we now have to calculate $S_{nn}(\omega)$.
The term $\langle \widetilde{R}_n^\dagger(\tau) R_n \rangle$ has been already calculated
in Eq.~\eqref{eq:RR}. The other term can be obtained through the same procedure, which leads to
\begin{multline}
  \langle R_n^\dagger \widetilde{R}_n(\tau) \rangle = \int dk \, |g_n(k)|^2 \\
  \times \! \left\{ e^{i\epsilon_n(k)\tau} \big[ 1 \! - \! \zeta f_n(\epsilon_n(k)) \big]
  + e^{-i\epsilon_n(k)\tau} f_n(\epsilon_n(k)) \right\} \!.
\end{multline}
Using the formula $\int_0^\infty e^{\pm i\epsilon\tau}d\tau = \pi \delta(\epsilon) \pm i \mathcal{P}[1/\epsilon]$, where $\mathcal{P}[1/\epsilon]$ stands for the Cauchy's principal value distribution, we see that
\begin{multline}
  S_{nn}(\omega) = \int dk \, |g_n(k)|^2 \times \\
  \left\{ \mathcal{P}\frac{1}{\omega + \epsilon_n(k)} f_n(\epsilon_n(k)) + \mathcal{P}\frac{1}{\omega - \epsilon_n(k)} \Big[ 1-\zeta f_n(\epsilon_n(k)) \Big] \right\} \!. \nonumber
\end{multline}
The quantity which enters in the Lamb-shift correction~\eqref{eq:middle_lambshift} is
$S_{nn}(\omega) - \zeta S_{nn}(-\omega)$.
Using the previous expression and the definition of the spectral density in Eq.~\eqref{eq:DOS},
we can then write
\begin{equation}
  \label{eq:new-lamb}
  H_{LS} = \sum_{k=1}^N \varphi_k b_k^\dagger b_k \,,
\end{equation}
where
\begin{equation}
  \label{eq:lambshift}
  \varphi_k = \, \frac{1}{\pi} \, \sum_{n=1}^{N_B} \Phi_{n,k}
     \left[ \mathcal{P} \int \frac{\mathcal{J}_n(\epsilon)}{\omega_k - \epsilon}d\epsilon + \zeta \mathcal{P}\int \frac{\mathcal{J}_n(\epsilon)}{\omega_k + \epsilon}d\epsilon \right] \, ,
\end{equation}
and this is the most general expression we can write without making assumptions about the spectral density.
Note that, if we assume our baths to have a very large bandwidth
with respect to the frequencies of the system,
\begin{equation}
  \mathcal{J}_n(\epsilon) \simeq \gamma > 0, \quad \forall \epsilon \geq 0 \,,
\end{equation}
the spectral density can be pulled out from the integrals,
which then become zero by means of the principal value sign.
Therefore, in such case one can safely assume $H_{LS} = 0$.
Otherwise, for the general case, one should evaluate the expression~\eqref{eq:lambshift}
according to the specific system-environment coupling model.
In any case, the Hermitian operator $H$ which appears in the master equation [Eq.~\eqref{eq:def_lambshift}]
is simply given by
\begin{equation}
   H = \sum_{k=1}^N \widetilde{\omega}_k b_k^\dagger b_k \, , \hspace{20pt}
   \text{with } \quad \widetilde{\omega}_k \equiv \omega_k + \varphi_k \,.
\end{equation}

Summarizing, we showed that it is possible to obtain a global LGKS master equation
for non-degenerate quadratic systems of the form in Eq.~\eqref{eq:LGKS}
where the Hermitian operator $H$
coincides with a possibly shifted version of the quadratic Hamiltonian
of the system~\eqref{eq:quadratic_hamiltonian}
and the dissipator $\mathcal{D}[\rho]$ is given by Eq.~\eqref{eq:dissipator}.
Notably, $\mathcal{D}[\rho]$ has the same form of the dissipator for the interaction
between a harmonic oscillator and a bath, as can be guessed from the fact that
a diagonalized Hamiltonian~\eqref{eq:diagonalized_hamiltonian} is equivalent
to a superposition of independent harmonic oscillators.
However, there are two important differences here: the Lindblad operators
are nonlocal (they are the normal modes of the system $b_k$) and the effective
coupling constants $\gamma_{n,k}$ explicitly depend on the BV matrices
via Eq.~\eqref{eq:coupling}.
The matrices $A$ and $B$ contain information about the spatial distribution of the
normal modes [see Eq.~\eqref{eq:BV}], therefore we expect this spatial form
to influence the couplings with the environment, as it should be.


\section{Steady state}
\label{sec:steadystate}

It is known that every LGKS master equation admits at least one steady state,
which is reached in the long-time limit $t \rightarrow \infty$. A more interesting point
concerns the uniqueness of such a state. In the literature, a number of theorems
have been proposed to characterize the conditions under which one can have a unique steady state,
however a conclusive statement on this subject is not simple to obtain~\cite{Nigro}.
Nevertheless, the Spohn theorem~\cite{Spohn} is sufficient to guarantee the uniqueness
of the steady state for the master equation constructed in Sec.~\ref{sec:rigorousME}.
Such theorem states that, if the set of Lindblad operators $\{L_i\}$ is self-adjoint
and its bicommutant $\{L_i\}''$ equals the entire operator space, then the steady state is unique.
We remind the reader that the commutant $\{L_i\}'$ is defined as the set of operators
which commute with all of the $L_i$, and the bicommutant $\{L_i\}''$ is simply
the commutant of the commutant.

In our case, $\{L_i\} = \{b_k\} \cup \{b_k^\dagger\}$. By definition, this is a self-adjoint set,
since for every $b_k$ the adjoint $b_k^\dagger$ always belongs to the set itself.
Moreover, due to the canonical rules, we know that there is no operator which simultaneously
commute with both $b_k$ and $b_k^\dagger$, except for the trivial one $\alpha I$, $\alpha \in \mathbb{C}$.
It follows that the commutant $\{L_i\}'$ is trivial and the bicommutant $\{L_i\}''$
equals the entire operator space. Due to the Spohn theorem we can therefore conclude that
the steady state is unique: the long-time dynamics is characterized by a well-defined relaxation process.
Obviously, in general, we have no reason to believe that this relaxation is of the thermal kind,
since the environment density operator $\rho_\mathcal{E}$ in Eq.~\eqref{eq:env_densityoperator}
is not characterized by a single temperature.
We then expect to deal with a nontrivial nonequilibrium steady state.

Let us see how to characterize this steady state through the observables.
We start from the adjoint version of our master equation, which can be written as
\begin{multline}
  \frac{dO_H(t)}{dt} = i \big\{ H, O_H(t) \big\}_- \\
  + \sum_{n,k} \gamma_{n,k} \Big[ (1 \! - \! \zeta f_n(\omega_k)) \left( 2b_k^\dagger O_H(t)b_k - \{ b_k^\dagger b_k, O_H(t) \}_+ \right) \\
   \phantom{+++} + f_n(\omega_k) \! \left( 2b_k O_H(t) b_k^\dagger - \big\{ b_k b_k^\dagger, O_H(t) \big\}_{\!+} \right) \! \Big] ,
\end{multline}
where $O_H(t)$ denotes the Heisenberg form of a Schr\"odinger observable $O$.
From this, it is possible to calculate the evolution of the expectation value
$\langle O_H(t) \rangle$. First of all, we consider two-point observables in quasiparticle operators.
The calculation is quite lengthy but straightforward, so here we just show the result:
\begin{subequations}
  \label{eq:evolution_SS_twopoint}
   \begin{eqnarray}
      \frac{d}{dt} \langle b_k^\dagger b_k \rangle &=& -2\sum_n \gamma_{n,k} \langle b_k^\dagger b_k \rangle + 2 \sum_n \gamma_{n,k} f_n(\omega_k) , \\
      \frac{d}{dt} \langle b_k^\dagger b_q \rangle &=& \Big[ i(\widetilde{\omega}_k - \widetilde{\omega}_q) - \sum_n (\gamma_{n,k} + \gamma_{n,q}) \Big] \langle b_k^\dagger b_q \rangle , \label{eq:evolution_SS_twopoint_b}\\
      \frac{d}{dt} \langle b_k^\dagger b_q^\dagger \rangle &=& \Big[ i(\widetilde{\omega}_k + \widetilde{\omega}_q) - \sum_n (\gamma_{n,k} + \gamma_{n,q}) \Big] \langle b_k^\dagger b_q^\dagger \rangle , \\
      \frac{d}{dt} \langle b_q b_k \rangle &=& \Big[ - \! i(\widetilde{\omega}_k + \widetilde{\omega}_q) \! - \!\! \sum_n (\gamma_{n,k} + \gamma_{n,q}) \Big] \! \langle b_q b_k \rangle , \qquad \;
   \end{eqnarray}
\end{subequations}
where in Eq.~\eqref{eq:evolution_SS_twopoint_b} it is assumed that $k\neq q$.
Note that all the four-point terms cancel each other, leaving us with only
two-point quantities. Every equation is closed by its own and each of them
can be easily integrated, leading to
\begin{subequations}
  \label{eq:SS_twopoint}
   \begin{eqnarray}
     \langle b_k^\dagger b_k \rangle(t) &=& \frac{\sum_n \gamma_{n,k} f_n(\omega_k)}{\sum_n \gamma_{n,k}}
     \left[ 1 - e^{-2\sum_n \gamma_{n,k}t} \right] \nonumber \\
     && \hspace*{1.5cm} + \langle b_k^\dagger b_k \rangle_0 \: e^{-2\sum_n \gamma_{n,k}t} \,, \\
      \langle b_k^\dagger b_q \rangle(t) &=& \langle b_k^\dagger b_q \rangle_0 \: e^{i( \widetilde{\omega}_k - \widetilde{\omega}_q)t - \sum_n (\gamma_{n,k}+\gamma_{n,q})t} \,, \\
      \langle b_k^\dagger b_q^\dagger \rangle(t) &=& \langle b_k^\dagger b_q^\dagger \rangle_0 \: e^{i( \widetilde{\omega}_k + \widetilde{\omega}_q)t - \sum_n (\gamma_{n,k}+\gamma_{n,q})t} \,, \\
      \langle b_q b_k \rangle(t) &=& \langle b_q b_k \rangle_0 \: e^{-i(\widetilde{\omega}_k + \widetilde{\omega}_q)t - \sum_n (\gamma_{n,k}+\gamma_{n,q})t} \,, \qquad
   \end{eqnarray}
\end{subequations}
where the subscript $\langle \cdot \rangle_0$ indicates the expectation value at time $t=0$.
If $\sum_n \gamma_{n,k} \neq 0$, the $t \rightarrow \infty$ limit leads to
\begin{equation}
  \label{eq:SS_solution1}
  \langle b_k^\dagger b_q \rangle_s = \delta_{kq} \frac{\sum_n \gamma_{n,k} f_n(\omega_k)}{\sum_n \gamma_{n,k}} \, ,
\quad
   \langle b_k^\dagger b_q^\dagger \rangle_s = \langle b_q b_k \rangle_s = 0 \, ,
\end{equation}
where the subscript $\langle \cdot \rangle_s$ refers to the expectation value on the steady state.
The only non-zero quantities are the diagonal occupations, which tend to the average
of the Fermi-Dirac (resp.~Bose-Einstein) distributions associated with the $N_B$ baths,
each weighted by the corresponding effective coupling $\gamma_{n,k}$.
This is an intuitive result, which confirms the nonequilibrium feature of the steady state.
Only in the case of perfectly identical baths $f_n(\omega_k) \equiv f(\omega_k)$,
we recover $\langle b_k^\dagger b_k \rangle_s = f(\omega_k)$,
independently of the details of the interaction.
Note that previous works have highlighted the appearance of steady-state
coherences in the presence of (quasi-)de\-ge\-ne\-ra\-cies~\cite{Harbola, Cattaneo-20, Dorn-21},
while here coherences are completely washed out: this fact is
known to be rooted in the full secular approximation that we assumed.

Eqs.~\eqref{eq:SS_twopoint} cease to be valid if, for some pair $(k,q)$,
it happens that $\gamma_{n,k} = \gamma_{n,q} = 0$ for all $n$.
In that case, we have to go back to Eqs.~\eqref{eq:evolution_SS_twopoint}
to understand that the new time-dependent solutions are
\begin{subequations}
  \label{eq:SS_solution2}
  \begin{eqnarray}
    \langle b_k^\dagger b_q \rangle(t) & = & \langle b_k^\dagger b_q \rangle_0 \: e^{i(\widetilde{\omega}_k - \widetilde{\omega}_q)t} \, , \\
    \langle b_k^\dagger b_q^\dagger \rangle(t) & = & \langle b_k^\dagger b_q^\dagger \rangle_0 \: e^{i(\widetilde{\omega}_k + \widetilde{\omega}_q)t} \, ,
  \end{eqnarray}
\end{subequations}
meaning that the expectation values remain the same as the initial ones,
apart from a phase factor of free evolution. This is consistent with the general idea
of open quantum system, since $\gamma_{n,k} = 0$ means that the $k$th mode is decoupled
from the $n$th bath and it only undergoes the unitary part of the evolution.

Equipped with Eqs.~\eqref{eq:SS_solution1} and~\eqref{eq:SS_solution2},
we can now calculate the steady-state correlation functions in real space
$C_{ij} \equiv \langle a_i^\dagger a_j \rangle_s$ and
$F_{ij} \equiv \langle a_i^\dagger a_j^\dagger \rangle_s$.
In order to calculate $C$, first note that
\begin{multline}
  \label{eq:aiaj}
  a_i^\dagger a_j = \sum_{k,q=1}^N \Big( A^*_{ik}A_{jq} b_k^\dagger b_q + B^*_{ik}B_{jq} b_k b_q^\dagger \\
  + A^*_{ik}B_{jq} b_k^\dagger b_q^\dagger + B^*_{ik}A_{jq} b_k b_q \Big) \, ,
\end{multline}
which, evaluated on the steady state, becomes
\begin{eqnarray}
   C_{ij} & = & \sum_k \left( A^*_{ik}A_{jk} \langle b_k^\dagger b_k \rangle_s + B^*_{ik}B_{jk} \langle b_k b_k^\dagger \rangle_s \right) \hspace*{1cm} \nonumber\\
   & = &\sum_k \left[ (A^*_{ik}A_{jk} \!- \!\zeta B^*_{ik}B_{jk}) \langle b_k^\dagger b_k \rangle_s + B^*_{ik}B_{jk} \right] \! ,
\end{eqnarray}
where, in the second equality, we used $b_k b_k^\dagger = 1- \zeta b_k^\dagger b_k$.
This result can be written in a compact matrix form as
\begin{equation}
  \label{eq:C}
   C = A^* \Theta A^T - \zeta B^* \Theta B^T + B^* B^T \, ,
\end{equation}
where we have defined the quasiparticle correlation matrix $\Theta_{kq} \equiv \langle b_k^\dagger b_q \rangle_s$.
The same procedure can be used to get
\begin{equation}
  \label{eq:F}
   F = A^* \Theta B^\dagger - \zeta B^* \Theta A^\dagger + B^* A^\dagger \, .
\end{equation}

We conclude this section by observing that higher-order observables can be calculated
in a similar fashion by means of the Wick theorem. For example, the four-point correlator
$G_{ij} \equiv \langle c_i^\dagger c_i c_j^\dagger c_j \rangle_s - \langle c_i^\dagger c_i \rangle_s \, \langle c_j^\dagger c_j \rangle_s$
is easily seen to be equal to $G_{ij} = F_{ij} F^*_{ji} - \zeta C_{ij} C_{ji} + \delta_{ij} C_{ii}$.


\section{Two-bath configuration}
\label{sec:twobath}

\begin{figure}
   \includegraphics[width=0.95\columnwidth]{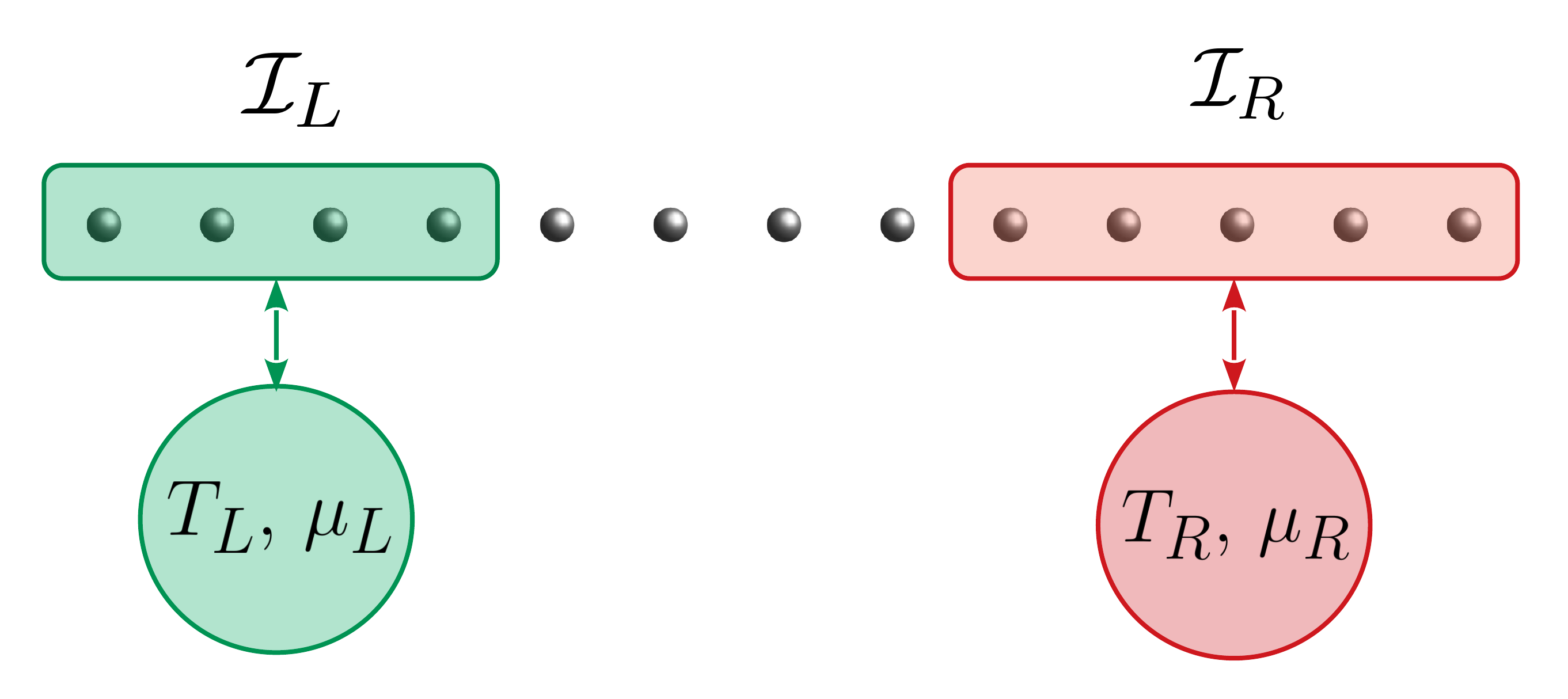}
   \caption{Sketch of the two-bath configuration proposed to study transport properties, in the particular case of one-dimensional lattices. The gray dots stand for the lattice sites, while the colored boxes stand for the regions of influence $\mathcal{I}_L$, $\mathcal{I}_R$ of the two baths.}
   \label{fig:transport}
\end{figure}

The presence of a system-environment interaction can be responsible for the appearance
of currents into the system, in general both of the electric kind (particle current)
and of thermal kind (energy and heat current). The standard way to deal with the analysis
of transport properties in quantum systems is based on the non-equilibrium Green's function
approach or the Landauer-B\"uttiker scattering matrix formalism~\cite{Datta}.
However, quite recently the master equation started to appear as well, as an interesting
alternative framework \cite{ME_transport, Jin-20}. In this section we follow this research line
and highlight the emergence of currents in the steady state of our master equation.

The derivation of Sec.~\ref{sec:rigorousME} is completely general and holds for any
interaction setting. In order to develop a framework which best describes the typical
experimental transport measurements, we specialize to the case of a two-bath configuration, $N_B=2$.
For the sake of clarity, in Fig.~\ref{fig:transport} we report a sketch of the situation
for the specific case of a one-dimensional lattice. However, we recall that our formalism
does not depend on the number of physical dimensions of the system.

If $N_B=2$, the dissipator defined in Eq.~\eqref{eq:dissipator} is composed of two terms,
coming from $n=1$ and $2$. To have a more appealing notation, we drop the use of the index $n$
and write the dissipator as $\mathcal{D}[\rho] = \mathcal{D}_L[\rho] + \mathcal{D}_R[\rho]$,
where the subscripts $_L$ and $_R$ respectively stand for ``left" and ``right",
referring to a hypothetical physical position of the two baths (see Fig.~\ref{fig:transport}).
We keep this notation in all the relevant quantities below.
For example, the effective coupling constants are denoted by $\gamma_{L,k}$ and $\gamma_{R,k}$,
and the quasiparticle correlation matrix is (for $\gamma_{L,k},\gamma_{R,k}\neq 0$) given by
\begin{equation}
  \label{eq:2bath_theta}
   \Theta_{kq} = \delta_{kq} \frac{\gamma_{L,k} f_L(\omega_k) + \gamma_{R,k} f_R(\omega_k)}{\gamma_{L,k} + \gamma_{R,k}} \, .
\end{equation}


\subsection{Particle and quasiparticle currents}

Let us start the analysis of the steady-state currents with the case of particle
transport (i.e., the electric current). In order to do that, we have to consider the evolution
equation for the total number of particles in the system, $\mathcal{N} = \sum_i a_i^\dagger a_i$.
The corresponding adjoint master equation reads
\begin{equation}
  \label{eq:2bath_particle_adjoint}
  \frac{d\mathcal{N}}{dt} = i \big\{ H, \mathcal{N} \big\}_-
  + \mathcal{D}_L^{(h)}[\mathcal{N}] + \mathcal{D}_R^{(h)}[\mathcal{N}] \, ,
\end{equation}
where $\mathcal{D}_L^{(h)}$ and $\mathcal{D}_R^{(h)}$ stand for the adjoint forms of the dissipators.
To calculate the expectation values, it is convenient to rewrite $\mathcal{N}$
using the normal modes $b_k$. This is done by simply taking the diagonal
of Eq.~\eqref{eq:aiaj}, $a^\dagger_i a_i$ and summing over the index $i$:
\begin{align}
  \mathcal{N} = &\sum_{k,q} \Big[ (A^\dagger A)_{kq} b_k^\dagger b_q + (B^\dagger B)_{kq} b_k b_q^\dagger \nonumber\\
    & \phantom{+} + (A^\dagger B)_{kq} b_k^\dagger b_q^\dagger + (B^\dagger A)_{kq} b_k b_q \Big]
  \equiv \sum_{k,q} \mathcal{N}_{kq}  \,.
\end{align}
After a long but straightforward calculation, one obtains
\begin{widetext}
\begin{subequations}
   \begin{eqnarray}
   & \big\{ H, \mathcal{N} \big\}_{\!-} \! = \sum_{k,q} \left\{ (\widetilde{\omega}_k - \widetilde{\omega}_q) \left[ (A^\dagger A)_{kq} b_k^\dagger b_q + \zeta (B^\dagger B)_{kq} b_q^\dagger b_k \right] + (\widetilde{\omega}_k + \widetilde{\omega}_q) \left[ (A^\dagger B)_{kq} b_k^\dagger b_q^\dagger + \zeta (B^\dagger A)_{kq} b_q b_k \right] \right\} , \quad \label{eq:2bath_particle_comm}\\
   & \mathcal{D}_L^{(h)}[\mathcal{N}] = 2\sum_{k} \gamma_{L,k} \left[ f_L(\omega_k) (A^\dagger A - \zeta B^\dagger B)_{kk} + (B^\dagger B)_{kk} \right] - \sum_{k,q} (\gamma_{L,k} + \gamma_{L,q}) \mathcal{N}_{kq} \, ,\\
   & \mathcal{D}_R^{(h)}[\mathcal{N}] = 2\sum_{k} \gamma_{R,k} \left[ f_R(\omega_k) (A^\dagger A - \zeta B^\dagger B)_{kk} + (B^\dagger B)_{kk} \right] - \sum_{k,q} (\gamma_{R,k} + \gamma_{R,q}) \mathcal{N}_{kq} \, .
   \end{eqnarray}
\end{subequations}
\end{widetext}

Now, when all of this is evaluated on the steady state of the master equation,
the only nonvanishing contributions are the diagonal normal ones.
It is easy to see that the commutator in Eq.~\eqref{eq:2bath_particle_comm} vanishes,
since the normal terms are multiplied by a factor $(\widetilde{\omega}_k - \widetilde{\omega}_q)$ which is zero for $k=q$.
Using the above definition for $\mathcal{N}_{kq}$, one can see that the left dissipator is
\begin{equation}
  \langle \mathcal{D}_L^{(h)}[\mathcal{N}] \rangle_s \! = \! 2 \sum_k \! \gamma_{L,k}
  ( A^\dagger A - \zeta B^\dagger B)_{kk} \big[ f_L(\omega_k) \! - \! \langle b_k^\dagger b_k \rangle_s \big] \! , \nonumber
\end{equation}
and the same expression is valid for $\mathcal{D}_R^{(h)}[\mathcal{N}]$,
after substituting $L \rightarrow R$. By definition $d\langle \mathcal{N} \rangle_s /dt = 0$,
so we can conclude that the adjoint master equation~\eqref{eq:2bath_particle_adjoint}
translates in the condition
\begin{multline}
  2\sum_k \gamma_{L,k} S_k \big[ f_L(\omega_k) - \langle b_k^\dagger b_k \rangle_s \big] \\
  = -2\sum_k \gamma_{R,k} S_k \big[ f_R(\omega_k) - \langle b_k^\dagger b_k \rangle_s \big] \, ,
  \label{eq:2bath_particle_condition}
\end{multline}
where we have introduced the factor
\begin{equation}
   S_k \equiv \big( A^\dagger A - \zeta B^\dagger B \big)_{kk} \, .
\end{equation}
To get the particle current from this, we note that the starting
point~\eqref{eq:2bath_particle_adjoint} has the shape of a quantum continuity equation~\cite{Mahan},
\begin{equation}
   \frac{d\langle \mathcal{N} \rangle}{dt} = J_\mathcal{N}^{(L)} + J_\mathcal{N}^{(R)} \, ,
\end{equation}
where $J_\mathcal{N}^{(L)}$ is the net particle current flowing from the left reservoir into the system,
while $J_\mathcal{N}^{(R)}$ is that flowing from the right reservoir into the system.
Evaluated on the steady state, $\langle \mathcal{N} \rangle$ does not change in time,
meaning that $J_\mathcal{N}^{(L)} = -J_\mathcal{N}^{(R)} \equiv J_\mathcal{N}$,
where $J_\mathcal{N}$ is the steady-state particle current.
However this is just the condition reported in Eq.~\eqref{eq:2bath_particle_condition} if one identifies
\begin{equation}
   J_\mathcal{N} \equiv 2\sum_{k} \gamma_{L,k} S_k \big[ f_L(\omega_k) - \langle b_k^\dagger b_k \rangle_s \big] \, .
\end{equation}
To conclude the derivation we just have to insert the value of $\langle b_k^\dagger b_k \rangle_s$.
If $\gamma_{L,k} = 0$ we have zero contribution from the $k$-th term of the sum,
so Eq.~\eqref{eq:2bath_theta} can be safely used to get
\begin{equation}
  \label{eq:2bath_partcurrent}
  J_\mathcal{N} = \sump_k \, \frac{2S_k \gamma_{L,k} \gamma_{R,k}}{\gamma_{L,k} + \gamma_{R,k}}
  \big[ f_L(\omega_k) - f_R(\omega_k) \big] \, ,
\end{equation}
where the prime sign in $\sump_k$ means that the sum runs only over those $k$
such that $\gamma_{L,k},\gamma_{R,k} \neq 0$. This expression has precisely the shape
of the Landauer-B\"uttiker formula, obtained with the scattering matrix approach:
the current is given by the difference between the Fermi-Dirac (or Bose-Einstein) distributions
of the two baths, weighted by a transfer factor which measures the easiness of the scattering
process~\cite{Datta}.

It is interesting to focus on the role of the quantity $S_k$, which appears in the transfer factor.
In order to do that, it is useful to construct a quantum continuity equation
for the total number of quasiparticles $\mathcal{N}_Q = \sum_k b_k^\dagger b_k$, instead of particles.
In such case, one has
\begin{equation}
  \frac{d\mathcal{N}_Q}{dt} = i \big\{ H, \mathcal{N}_Q \big\}_-
  + \mathcal{D}_L^{(h)}[\mathcal{N}_Q] + \mathcal{D}_R^{(h)}[\mathcal{N}_Q] \, ,
\end{equation}
which is a greatly simplified situation, with respect to the previous one,
since we do not need a BV transformation here.
As a matter of fact, the steady-state quasiparticle current can be directly seen to be
\begin{equation}
  J_{\mathcal{N}_Q} = \sump_k \, \frac{2\gamma_{L,k}\gamma_{R,k}}{\gamma_{L,k} + \gamma_{R,k}}
  \big[ f_L(\omega_k) - f_R(\omega_k) \big] \, ,
\end{equation}
which is the same as the particle current in Eq.~\eqref{eq:2bath_partcurrent},
but without $S_k$. Note that here the transfer factor is the same
as the well-known one for a set of independent ballistic channels~\cite{Datta}.

Remember that if $H_\mathcal{S}$ does not contain
anomalous terms, we can arrange the BV transformation in such a way to make $\mathcal{N} = \mathcal{N}_Q$.
Obviously, in this case the two kinds of currents coincide $J_\mathcal{N} = J_{\mathcal{N}_Q}$ and,
indeed, this is confirmed by the value $S_k \equiv 1$ (since $B=0$ and $A$ is unitary).
In the general case, $S_k \neq 1$ and the particle current has a different transfer factor.
For this reason we propose to call $S_k$ anomaly factor, since it emerges because of
the presence of anomalous terms in the system Hamiltonian $H_\mathcal{S}$.
The presence of $S_k$ makes the particle transfer factor deviate from
the standard form, thus it is a potentially crucial quantity of our theory.
A more thorough study of the anomaly factor and its effects on the transport properties
is left for subsequent works.


\subsection{Energy current}

The steady-state quasiparticle current $J_{\mathcal{N}_Q}$ has no actual experimental meaning,
since the particles are the physical entities which actually move along the system.
However, $J_{\mathcal{N}_Q}$ is important for the study of energy-related phenomena,
since the quasiparticles are the mathematical objects which are responsible for thermal transport,
if present. Indeed, a quantum continuity equation for the
free Hamiltonian of the system $H_\mathcal{S}$ can be constructed from the adjoint master equation
\begin{equation}
   \frac{dH_{\mathcal S}}{dt} = \mathcal{D}_L^{(h)}[H_\mathcal{S}] + \mathcal{D}_R^{(h)}[H_\mathcal{S}] \, ,
\end{equation}
where we have used $\big\{ H , H_\mathcal{S} \big\}_- = 0$.
From this, it is easy to see that the steady-state energy current is
\begin{equation}
  J_E = \sump_k \, \frac{2 \, \omega_k \gamma_{L,k} \gamma_{R,k}}{\gamma_{L,k} + \gamma_{R,k}}
  \big[ f_L(\omega_k) - f_R(\omega_k) \big] \, .
\end{equation}
This expression is the same as $J_{\mathcal{N}_Q}$, apart from the appearance of $\omega_k$
in the transfer factor, for dimensionality reasons.
This looks like the natural generalization of previous results obtained
in the literature in simpler contexts (see, e.g., Ref.~\cite{Hofer-17}).


\subsection{Heat current and consistency with the thermodynamics}

To create an imbalance $f_L \neq f_R$ which can generate the currents, we have control of both
the chemical potentials $\mu_n$ and the temperatures $T_n$ of the two baths.
Note that a particle current can be created even with a temperature imbalance only and,
in the same way, an energy current with an electrical imbalance only.
These are the so-called thermoelectric effects, which are known to exist
in many-body systems~\cite{Mahan, Benenti-rev-17}.

In our context it is worth elaborating on this issue, because it is an easy way
to test the detailed balance condition, which is known to be valid for any LGKS master equation
with an equilibrium steady state. The master equation constructed in this work
has an equilibrium steady state in the case of identical baths,
where $\langle b_k^\dagger b_k \rangle_s = f(\omega_k)$.
The link between the detailed balance condition (characterizing an equilibrium situation)
and the thermoelectric effects (characterizing a non-equilibrium situation)
is the Onsager relation~\cite{Onsager}.
To properly define it, let us suppose to have infinitesimal imbalances
$\mu_{L/R} = \mu \pm \Delta \mu/2$ and $T_{L/R} = T \pm \Delta T/2$.
In standard many-body theory, this is sufficient to define the steady-state heat current
as $J_\mathcal{Q} \equiv J_E - \mu J_{\mathcal{N}_Q}$~\cite{Mahan}.
The thermoelectric transport coefficients are then defined by the following:
\begin{equation}
  \label{eq:onsager_matrix}
   \left(
   \begin{array}{c}
      J_{\mathcal{N}_Q} \\ J_\mathcal{Q}
   \end{array}
   \right) = \left(
   \begin{array}{cc}
      \ell_{11} & \ell_{12} \\ \ell_{21} & \ell_{22}
   \end{array}
   \right) \left(
   \begin{array}{c}
      \Delta\mu / T \\ \Delta T/T^2
   \end{array}
   \right) \, ,
\end{equation}
where $\{\ell_{ij}\}$ is the Onsager matrix. The Onsager relation tells us that,
if in the equilibrium situation the system obeys a detailed balance condition,
then the Onsager matrix is symmetric, that is, $\ell_{12} = \ell_{21}$.

We expect the Onsager relation to be valid in the context of transport through quadratic systems.
In order to verify that, let us start by noticing that we can perform the expansion
\begin{equation}
   f_L(\omega_k) - f_R(\omega_k) \simeq \frac{\partial f(\omega_k)}{\partial T} \Delta T + \frac{\partial f(\omega_k)}{\partial \mu} \Delta \mu \, ,
\end{equation}
where $f(\omega) \equiv [\zeta + e^{(\omega - \mu)/T}]^{-1}$. With this formula, we can rewrite the currents as
\begin{subequations}
  \label{eq:onsager_currents}
   \begin{eqnarray}
      J_{\mathcal{N}_Q} & = & \frac{\partial F_N}{\partial T} \Delta T + \frac{\partial F_N}{\partial \mu} \Delta \mu \, , \\
      J_E & = & \frac{\partial F_E}{\partial T} \Delta T + \frac{\partial F_E}{\partial \mu} \Delta \mu \, ,
   \end{eqnarray}
\end{subequations}
after introducing
\begin{eqnarray}
  F_N & \equiv & \sump_k \, \frac{2\gamma_{L,k}\gamma_{R,k}}{\gamma_{L,k} + \gamma_{R,k}} f(\omega_k) \, ,
  \nonumber \\
  F_E & \equiv & \sump_k \frac{2\gamma_{L,k} \gamma_{R,k}}{\gamma_{L,k} + \gamma_{R,k}} \omega_k f(\omega_k) \, .
  \nonumber
\end{eqnarray}
We note that the same kind of expansion was performed in Ref.~\cite{Santos} for the infinite bosonic
tight-binding chain. Here we have shown that it is actually valid for generic finite quadratic
Hamiltonians, provided the correct expressions for $F_N$ and $F_E$ are used.

Comparing Eqs.~\eqref{eq:onsager_currents} with the definition of the Onsager matrix
in Eq.~\eqref{eq:onsager_matrix}, we immediately obtain the transport coefficients
in terms of derivatives of the generating functions $F_N$, $F_E$ as
\begin{subequations}
   \begin{align}
      \ell_{11} &= T\frac{\partial F_N}{\partial \mu} \, , & \ell_{22} &= T^2 \left( \frac{\partial F_E}{\partial T} - \mu \frac{\partial F_N}{\partial T} \right) \, , \\
      \ell_{12} &= T^2 \frac{\partial F_N}{\partial T} \, , & \ell_{21} &= T \left( \frac{\partial F_E}{\partial \mu} - \mu \frac{\partial F_N}{\partial \mu} \right) \, .
   \end{align}
\end{subequations}
Using the fact that
\begin{equation}
   T \frac{\partial f(\omega_k)}{\partial T} = (\omega_k - \mu) \frac{\partial f(\omega_k)}{\partial \mu} \, ,
\end{equation}
we finally see that
\begin{align}
  \ell_{12} & = T^2 \frac{\partial F_N}{\partial T} = T^2 \sump_k \, \frac{2\gamma_{L,k}\gamma_{R,k}}{\gamma_{L,k} + \gamma_{R,k}} \frac{\partial f(\omega_k)}{\partial T} \nonumber \\
  & = T \sump_k \, \frac{2\gamma_{L,k}\gamma_{R,k}}{\gamma_{L,k} + \gamma_{R,k}} (\omega_k - \mu) \frac{\partial f(\omega_k)}{\partial \mu} \nonumber \\
  & = T \sump_k \, \frac{2\gamma_{L,k}\gamma_{R,k}}{\gamma_{L,k} + \gamma_{R,k}} \left[ \omega_k \frac{\partial f(\omega_k)}{\partial \mu} - \mu \frac{\partial f(\omega_k)}{\partial \mu} \right] \nonumber \\
  & = T \left( \frac{\partial F_E}{\partial \mu} - \mu \frac{\partial F_N}{\partial \mu} \right) = \ell_{21} \, .
\end{align}
which is precisely the Onsager relation. This is an important conceptual result,
since it shows that our approach to quantum transport through quadratic systems
permits us to reobtain long-standing results in the context of nonequilibrium thermodynamics.


\section{Conclusions and Outlook}
\label{sec:conclusions}

We have discussed how to derive a wide class of Lindblad-type master equations
for generic quadratic quantum many-body systems, where the interaction with a set
of independent thermal baths is properly taken into account.
Having relaxed the commonly employed local approximation for the system-environment
coupling, the only limitation of our treatment resides in the Born-Markov and secular hypotheses.
In particular, our approach reconciles all the thermodynamic
inconsistencies that may emerge in ordinary many-body approaches where the Lindblad
jump operators act locally in the physical space of the system.

The resulting nonlocal master equations can be easily solved to obtain time-dependent
correlation functions, using an amount of resources analogous to that for a local
approach (i.e., scaling polynomially in the system size).
This paves the way for the study of nonequilibrium Markovian dynamics of quadratic
quantum systems, in situations where nonlocality cannot be overlooked (as in
solid-state~\cite{Franz-13} or hybrid photonic devices~\cite{{Blais-11}}), with
up to a few thousand sites. Within our framework, interactions should be
treated at a mean-field level.

Several many-body aspects are worth being investigated, including the emergence
of dissipation-driven transitions, the role of a time-dependent external driving,
as well as the robustness of quantum transport phenomena or of topological states
to the presence of unitary and/or dissipative disorder.
It would also be tempting to study thermodynamic processes at the nanoscale
in the context of heat engines~\cite{Benenti-rev-17, Nori-07},
where the presence of critical modes may affect the heat-to-work
efficiency~\cite{Campisi-16}.

Finally, investigations in the context of quadratic systems could be pushed
further to develop more accurate master equations,
going beyond the Born-Markov and secular approximations.
In fact, we believe this is a mandatory step to address the description
of more realistic situations.


\appendix

\section{Systems with a zero-energy mode}
\label{appendix}

In Subsec.~\ref{subsec:dissipator} we provided the calculation of a rigorous nonlocal
dissipator [reported in Eq.~\eqref{eq:dissipator}], valid as long as the system
of interest $\mathcal{S}$ is non-degenerate and does not support a zero-energy mode.
In this appendix we relax the second hypothesis: we still have non-degenerate
eigenenergies, but a zero mode is now present.
According to the discussion of Sec.~\ref{sec:quadratic},
here we limit our analysis to the case of a fermionic zero mode ($\zeta = 1$).

The starting point is Eq.~\eqref{eq:OnOn}, from which one clearly obtains additional terms
with respect to Eq.~\eqref{eq:OnOn2}. As a convention, let us indicate with $k=0$ the index
associated with the zero-energy mode $\omega_0 = 0$.
The equivalent of the expression~\eqref{eq:OnOn2} is
\begin{align}
  \sum_{p,s;\, k} & \left[ \phi_{pk} \, \phi^*_{sk} \, \delta_{\omega,\omega_k} \, b_k \rho b_k^\dagger +
    \phi^*_{pk} \, \phi_{sk} \, \delta_{\omega,-\omega_k} \, b_k^\dagger \rho b_k \right] \nonumber \\
  + \sum_{p,s} & \: \delta_{\omega,0} \left[ \phi_{p0} \, \phi_{s0} \, b_0 \rho b_0
    + \phi^*_{p0} \, \phi^*_{s0} \, b_0^\dagger \rho b_0^\dagger \right] \,.
\end{align}
The same thing can be done for the other terms in Eq.~\eqref{eq:our_dissipator_diagonal}
and the result is
\begin{equation}
   \mathcal{D}[\rho] = \mathcal{D}^{(\rm st)}[\rho] + \sum_n 2 \Gamma_{nn}(0)
   \left( \Psi_{n,0} b_0 \rho b_0 + \Psi^*_{n,0} b_0^\dagger \rho b_0^\dagger \right)
\end{equation}
where $\mathcal{D}^{(\rm st)}[\rho]$ is the standard dissipator of
Eq.~\eqref{eq:middle_dissipator} and
\begin{equation}\label{eq:Psi}
  \Psi_{n,k} \equiv \sum_{p,s\in\mathcal{I}_n} w_{p,n} \, w_{s,n} \, \phi_{pk} \, \phi_{sk}
  = \Big( \sum_{p\in\mathcal{I}_n} w_{p,n} \, \phi_{pk} \Big)^2 \,.
\end{equation}
Here $\mathcal{D}[\rho]$ can be written in the LGKS form by extracting the term $k=0$ from
$\mathcal{D}^{(\rm st)}[\rho]$ and putting it in the additional term. If we call
$\mathcal{D}^{(\rm st)}_{>0}[\rho]$ the term $\mathcal{D}^{(\rm st)}[\rho]$ deprived of
the $k=0$ term, it is possible to see that
\begin{multline}
   \mathcal{D}[\rho] = \mathcal{D}^{(\rm st)}_{>0}[\rho] + \sum_n \Gamma_{nn}(0) \times \\
   \times \Big[ 2\Big( \sqrt{\Phi_{n,0}} b_0 + \frac{\Psi^*_{n,0}}{\sqrt{\Phi_{n,0}}} b_0^\dagger \Big) \rho \Big( \frac{\Psi_{n,0}}{\sqrt{\Phi_{n,0}}} b_0 + \sqrt{\Phi_{n,0}} b_0^\dagger \Big) \\
   - \Big\{ \Big( \frac{\Psi_{n,0}}{\sqrt{\Phi_{n,0}}} b_0 + \sqrt{\Phi_{n,0}} b_0^\dagger \Big) \Big( \sqrt{\Phi_{n,0}} b_0 + \frac{\Psi^*_{n,0}}{\sqrt{\Phi_{n,0}}} b_0^\dagger \Big), \rho \Big\}_+ \Big]
\end{multline}
where, without loss of generality, we have assumed $\Phi_{n,0}, \Psi_{n,0} \neq 0$ and we used that
$|\Psi_{n,0}|^2 = \Phi^2_{n,0}$.

To obtain a simpler expression, we will limit ourselves to the case of
a real system Hamiltonian $H_\mathcal{S}$, where $\Phi \equiv \Psi$.
The complex case can be handled in a similar manner.
Using Eq.~\eqref{eq:gamma} to express $\Gamma_{nn}(\omega)$ we finally reach
\begin{subequations}
   \begin{equation}
      \mathcal{D}[\rho] = \mathcal{D}^{(\rm st)}_{>0}[\rho] + \mathcal{D}_0[\rho] \,,
   \end{equation}
   where
   \begin{equation}
     \mathcal{D}_0[\rho] = \Delta \Big[ 2 \big( b_0 + b_0^\dagger \big) \rho \big( b_0 + b_0^\dagger \big)
       - \big\{ \big( b_0 + b_0^\dagger \big)^2, \rho \big\}_+ \Big]
   \end{equation}
   and we have introduced the constant
   \begin{equation}
      \Delta \equiv \sum_n \mathcal{J}_n(0) \, \Phi_{n,0} = \sum_n \gamma_{n,0} \,.
   \end{equation}
\end{subequations}

As before, to conclude the derivation of the master equation, we should check for
the Lamb-shift correction. With the same procedure as before, it is easy to see that
\begin{equation}
   H_{LS} = H_{LS}^{(\rm st)} \,,
\end{equation}
where $H_{LS}^{(\rm st)}$ is the expression in Eq.~\eqref{eq:middle_lambshift_pr}.
Therefore, the shape of the Lamb-shift correction is not influenced by the presence
of the zero mode (in the fermionic case).

The set of Lindblad operators is now $\{ L_i \} = \{ b_k \} \cup \{ b_k^\dagger \}
\cup \{ b_0 + b_0^\dagger \}$. This is still a self-adjoint set with a trivial commutant,
so the Spohn theorem is valid~\cite{Spohn} and the steady state of the dynamics is unique also in this case.
As before, it can be characterized by two-point observables in the quasiparticle operators
$\langle b_k^\dagger b_q \rangle$, $\langle b_k b_q \rangle$, $\langle b_k^\dagger b_q^\dagger \rangle$.
If $k,q \neq 0$ it is easy to see that the evolution equations reduce to the ones reported in
Eqs.~\eqref{eq:evolution_SS_twopoint}, therefore the non-zero components of the quasiparticle
correlation matrix $\Theta$ are unaffected by the presence of the zero mode.
The other relevant equations turn out to be
\begin{eqnarray}
  \frac{d}{dt} b_0^\dagger b_0 &=& -4\Delta b_0^\dagger b_0 + 2\Delta \,, \nonumber \\
  \frac{d}{dt} b_0^\dagger b_q &=& \Big[ i (\widetilde{\omega}_0 - \widetilde{\omega}_q) - \! 2\Delta - \!\sum_n \! \gamma_{n,q} \Big] b_0^\dagger b_q - 2\Delta b_0 b_q , \nonumber \\
  \frac{d}{dt} b_0 b_q &=& \Big[ - i (\widetilde{\omega}_0 + \widetilde{\omega}_q) - \! 2\Delta - \!\sum_n \! \gamma_{n,q} \Big] b_0 b_q - 2\Delta b_0^\dagger b_q . \qquad \; \nonumber
\end{eqnarray}
where it is implicitly assumed that $q \neq 0$. All the other equations can be obtained from these by taking their adjoints.
Notice that
now we have obtained a coupled system of differential equations.
Nevertheless, the above system is linear and can be easily solved to obtain
\begin{subequations}
   \begin{eqnarray}
      \langle b_0^\dagger b_0 \rangle(t) &=& \Big[ \langle b_0^\dagger b_0 \rangle_0 - \tfrac{1}{2} \Big] e^{-4\Delta t} + \tfrac{1}{2}, \phantom{++++++++}\\
      \langle b_0^\dagger b_q \rangle(t) &=& \tfrac{1}{2} e^{i (\widetilde{\omega}_0 - \widetilde{\omega}_q) t - \sum_n \gamma_{n,q}t} \times \nonumber\\
      && \hspace*{-1.1cm} \times \!\left[ \langle b_0^\dagger b_q \rangle_0 \left( e^{-4\Delta t} \!+ \!1 \right) \!+ \! \langle b_0 b_q \rangle_0 \left( e^{-4\Delta t} \!-\! 1 \right) \right] \!, \\
      \langle b_0 b_q \rangle(t) &=& \tfrac{1}{2} e^{-i (\widetilde{\omega}_0 + \widetilde{\omega}_q) t - \sum_n \gamma_{n,q}t} \times \nonumber\\
      && \hspace*{-1.1cm} \times \!\left[ \langle b_0^\dagger b_q \rangle_0 \left( e^{-4\Delta t} \!- \!1 \right) \!+ \! \langle b_0 b_q \rangle_0 \left( e^{-4\Delta t} \!+\! 1 \right) \right] \!.
   \end{eqnarray}
\end{subequations}
For $t \rightarrow \infty$, the first relation indicates that $\langle b_0^\dagger b_0 \rangle_s = 1/2$,
independently of the interaction setting. The other quantities decay to zero,
provided $\sum_n \gamma_{n,q} \neq 0$ [if this is not the case, they display an oscillatory behavior,
  analogously to Eq.~\eqref{eq:SS_solution2}]. We conclude that the
quasiparticle correlation matrix is now given by
\begin{equation}
   \Theta_{kq} = \makeatletter \bBigg@{5}\{ \makeatother
   \begin{array}{ll}
      \delta_{kq} \dfrac{\sum_n \gamma_{n,k}f_n(\omega_k)}{\sum_n \gamma_{n,k}} & \text{if } k,q \neq 0 \,,\\ \vspace*{2mm}
      1/2 & \text{if } k = q = 0 \,, \\ \vspace*{2mm}
      0 & \text{otherwise} \,.
   \end{array}
\end{equation}
However, the expressions for the correlation functions in real space
[Eqs.~\eqref{eq:C} and \eqref{eq:F}] remain unaffected.

The same kind of calculation can be performed to study the steady-state currents
in a minimal two-bath configuration. For example, the adjoint master equation for
the total number of particles $\mathcal{N}$ acquires an additional term
$\mathcal{D}_0^{(h)}[\mathcal{N}] = \mathcal{D}_0[\mathcal{N}]$
with respect to Eq.~\eqref{eq:2bath_particle_adjoint}, which is equal to
\begin{eqnarray}
   &&\mathcal{D}_0[\mathcal{N}] = 2\Delta \sum_k \Big\{ \big[ (B^\dagger B + B^\dagger A)_{k0} - (A^\dagger A + B^\dagger A)_{0k} \big] \nonumber\\
   &&\times (b_0^\dagger b_k + b_0 b_k) - \big[ (A^\dagger A + A^\dagger B)_{k0} - (B^\dagger B + A^\dagger B)_{0k} \big] \nonumber\\
   &&\times (b_k^\dagger b_0 + b_k^\dagger b_0^\dagger) \Big\} + 2\Delta (A^\dagger A - B^\dagger B)_{00} \,.
\end{eqnarray}

When evaluated for the steady state, only the terms with $\langle b_0^\dagger b_0 \rangle_s$
remain and therefore we can immediately see that $\langle \mathcal{D}_0[\mathcal{N}] \rangle_s = 0$.
This means that the particle current in Eq.~\eqref{eq:2bath_partcurrent} is not affected by the
presence of the zero-energy mode, provided the term $k=0$ is excluded from the sum.
The same reasoning can be applied to the quasiparticle current and the energy current,
where the additional term to the adjoint master equation turns out to be
\begin{equation}
   \mathcal{D}_0[\mathcal{N}_Q] = 2 \Delta \big( 1-2b_0^\dagger b_0 \big)
\end{equation}
and then $\langle \mathcal{D}_0[\mathcal{N}_Q] \rangle_s = 0$, as before.


\section{Systems with degenerate eigenenergies}
\label{appendixB}

The analysis performed in Subsec.~\ref{subsec:dissipator} is valid for non-degenerate systems,
so that $\omega_k = \omega_q$ only if $k=q$. In this appendix we briefly discuss how it is
possible to include the presence of degenerate eigenenergies into our formalism.

Let us start from Eq.~\eqref{eq:OnOn} and suppose that the system $\mathcal{S}$ possesses $M$
different energy eigenspaces, labeled by an index $\lambda = 1,\ldots,M$. We indicate with
$\mathcal{A}_\lambda$ the set of normal-modes indexes associated with the $\lambda$th
eigenspace, with eigenvalue $\omega_\lambda$. For the moment, let us also suppose for the sake
of simplicity that there are no zero-energy modes, i.e. $\omega_\lambda \neq 0$, for all $\lambda$.
Then:
\begin{eqnarray}
   O_n(\omega) \rho \, O_n^\dagger(\omega)  =  \sum_{p,s\in\mathcal{I}_n} \, \sum_{\lambda=1}^M \, \sum_{u,v\in\mathcal{A}_\lambda} w_{p,n} w^*_{s,n} \times \nonumber \\
    \times \Big[ \delta_{\omega,\omega_\lambda} \phi_{pu} \, \phi^*_{sv} \, b_u \rho b_v^\dagger   + \delta_{\omega,-\omega_\lambda} \phi^*_{pu} \, \phi_{sv} \, b_u^\dagger \rho b_v \Big] \,.
   \label{eq:Bbeginning}
\end{eqnarray}
The same thing can be done for the other terms of Eq.~\eqref{eq:our_dissipator_diagonal} and the
result for the dissipator is:
\begin{eqnarray}
   \mathcal{D}[\rho] & = & \sum_{n;\, \lambda} \sum_{u,v\in\mathcal{A}_\lambda} \!\! \Big[ \Phi^{(n,\lambda)}_{uv} \Gamma_{nn}(\omega_\lambda) \Big( 2b_u \rho b_v^\dagger - \big\{ b_v^\dagger b_u, \rho \big\}_{\!+} \Big) \nonumber\\
     && + \Phi^{(n,\lambda)}_{vu} \, \Gamma_{nn}(-\omega_\lambda) \Big( 2b_u^\dagger \rho b_v - \big\{ b_v b_u^\dagger, \rho \big\}_{\!+} \Big) \Big] \,,
   \label{eq:dissipator_degenerate}
\end{eqnarray}
where
\begin{equation}
   \Phi^{(n,\lambda)}_{uv} \equiv \sum_{p,s\in\mathcal{I}_n} w_{p,n} \, w^*_{s,n} \, \phi_{pu} \, \phi^*_{sv}
\end{equation}
are the elements of a rank-one Hermitian matrix.
Notice that this quantity constitutes the generalization to the degenerate case of the quantity
$\Phi_{n,k}$ defined in Eq.~\eqref{eq:Phi}. The index $\lambda$ here is needed to indicate
that $u,v\in\mathcal{A}_\lambda$, therefore it fixes the dimension of the matrix.

Once we use Eq.~\eqref{eq:gamma} to write $\Gamma_{nn}(\omega)$, Eq.~\eqref{eq:dissipator_degenerate}
is already a dissipator in the LGKS form, which can eventually be studied.
Notice that it is not diagonal anymore and the Spohn theorem then ceases to be valid.
This means that the presence of degeneracies in $H_\mathcal{S}$ can make the system develop multiple steady-state solutions.

It is also worth pointing out that the inequality $\Phi_{n,k} \geq 0$ translates here
in a positive semi-definiteness requirement for the matrix $\Phi^{(n,\lambda)}$.
In order to see that, note that for
fixed $(n,\lambda)$, the matrix $\Phi^{(n,\lambda)}$ is Hermitian, hence it is diagonalized by a unitary
matrix $U^{(n,\lambda)}$. Let us then write
\begin{equation}\label{eq:Phitilde}
   U^{(n,\lambda)\dagger} \, \Phi^{(n,\lambda)} \, U^{(n,\lambda)} \equiv \widetilde{\Phi}^{(n,\lambda)} \,,
\end{equation}
where $\widetilde{\Phi}^{(n,\lambda)}$ is a real diagonal matrix.
For its elements, we can see that
\begin{eqnarray}
   \widetilde{\Phi}^{(n,\lambda)}_{ww}  =  \sum_{p,s\in\mathcal{I}_n} \sum_{u,v\in\mathcal{A}_\lambda} U^{(n,\lambda)*}_{uw} w_{p,n} \phi_{pu} \phi^*_{sv} w_{s,n}^* U^{(n,\lambda)}_{vw} \nonumber \\
    =  \Big( \sum_{p\in\mathcal{I}_n} \sum_{u\in\mathcal{A}_\lambda} U^{(n,\lambda)*}_{uw} w_{p,n} \phi_{pu} \Big) \Big( \sum_{s\in\mathcal{I}_n} \sum_{v\in\mathcal{A}_\lambda} U^{(n,\lambda)}_{vw} w_{s,n}^* \phi^*_{sv} \Big) \nonumber \\
    =  \Big| \sum_{p\in\mathcal{I}_n} \sum_{u\in\mathcal{A}_\lambda} U^{(n,\lambda)*}_{uw} w_{p,n} \phi_{pu} \Big|^2 \geq 0 \,. \nonumber
\end{eqnarray}
Notice also that the same procedure can be used to calculate the Lamb-shift correction,
which turns out to be the following matrix-like generalization of Eq.~\eqref{eq:new-lamb}:
\begin{equation}
   H_{LS} = \sum_{\lambda=1}^M \sum_{u,v\in\mathcal{A}_\lambda} \varphi_{uv} b_u^\dagger b_v \, ,
\end{equation}
where
\begin{equation}
   \varphi_{uv} \equiv \sum_{n=1}^{N_B} \Phi^{(n,\lambda)}_{vu} \Big[ S_{nn}(\omega_\lambda) - \zeta S_{nn}(-\omega_\lambda) \Big] \, .
\end{equation}

We conclude by considering the case in which we also relax the constraint on the absence of zero-energy modes
for fermionic systems.
Let us indicate with $\mathcal{A}_0$ the set of normal-modes indexes associated with
the eigenspace with $\omega_0 = 0$. Equation~\eqref{eq:Bbeginning} then clearly acquires an additional term given by
\begin{equation}
  \sum_{p,s\in\mathcal{I}_n} \sum_{u,v\in\mathcal{A}_0} \delta_{\omega,0} \, w_{p,n} \, w_{s,n}
  \Big[ \phi_{pu}\phi_{sv} \, b_u \rho b_v + \phi^*_{pu}\phi^*_{sv} \, b_u^\dagger \rho b_v^\dagger \Big] \,. \nonumber
\end{equation}
The dissipator becomes
\begin{eqnarray}
   \mathcal{D}[\rho] & = & \mathcal{D}^{(\rm st)}[\rho] + \sum_{n=1}^{N_B} \sum_{u,v\in\mathcal{A}_0} \Gamma_{nn}(0) \nonumber \\
   && \times \Big[ \Psi^{(n,0)}_{uv} \Big( 2b_u \rho b_v - \big\{ b_v b_u, \rho \big\}_+ \Big) \nonumber \\
     && + \Psi^{(n,0)*}_{uv} \Big( 2b_u^\dagger \rho b_v^\dagger - \big\{ b_v^\dagger b_u^\dagger, \rho \big\}_+ \Big) \Big] \,,
\end{eqnarray}
where $\mathcal{D}^{(\rm st)}[\rho]$ is the dissipator in Eq.~\eqref{eq:dissipator_degenerate} and
\begin{equation}
  \Psi^{(n,\lambda)}_{uv} \equiv \sum_{p,s\in\mathcal{I}_n} w_{p,n} \, w_{s,n} \, \phi_{pu} \, \phi_{sv}
\end{equation}
is the generalization to the degenerate case of the quantity $\Psi_{n,k}$ defined back in
Eq.~\eqref{eq:Psi}. As done in App.~\ref{appendix}, at this point it is sufficient to extract
the term with $\lambda = 0$ from $\mathcal{D}^{(\rm st)}[\rho]$ and put it in the additional term
to obtain a LGKS dissipator.


\end{document}